\documentclass[preprint,rotating.sty]{aastex}
\begin{document}
\title{Cataclysmic Variables from SDSS II. The Second Year \footnote{Based on 
observations obtained with the Sloan Digital Sky Survey and with the
 Apache Point
Observatory (APO) 3.5m telescope, which are owned and operated by the
Astrophysical Research Consortium (ARC)}}

\author{Paula Szkody\altaffilmark{2}, 
Oliver Fraser\altaffilmark{2}, Nicole Silvestri\altaffilmark{2},
Arne Henden\altaffilmark{3,4},
Scott F. Anderson\altaffilmark{2}, 
James Frith\altaffilmark{2},
Brandon Lawton\altaffilmark{2},
Ethan Owens\altaffilmark{2},
Sean Raymond\altaffilmark{2}, 
Gary Schmidt\altaffilmark{5},
Michael Wolfe\altaffilmark{2},
John Bochanski\altaffilmark{2}, Kevin Covey\altaffilmark{2},
Hugh Harris\altaffilmark{3},
Suzanne Hawley\altaffilmark{2},
Gillian R. Knapp\altaffilmark{6}, 
Bruce Margon\altaffilmark{7},  
Wolfgang Voges\altaffilmark{8}, 
Lucianne Walkowicz\altaffilmark{2},
J. Brinkmann\altaffilmark{9},
D. Q. Lamb\altaffilmark{10} } 

\altaffiltext{2}{Department of Astronomy, University of Washington, Box 351580,
Seattle, WA 98195}
\altaffiltext{3}{US Naval Observatory, Flagstaff Station, P. O. Box 1149,
Flagstaff, AZ 86002-1149}
\altaffiltext{4}{Universities Space Research Association}
\altaffiltext{5}{The University of Arizona, Steward Observatory, Tucson, AZ 85721}
\altaffiltext{6}{Princeton University Observatory, Princeton, NJ 08544}
\altaffiltext{7}{Space Telescope Science Institute, Baltimore, MD 21218}
\altaffiltext{8}{Max-Planck-Institute f\"ur extraterrestrische Physik,
Geissenbachstr. 1, D-85741 Garching, Germany}
\altaffiltext{9}{Apache Point Observatory, P.O. Box 59, Sunspot, NM 88349-0059}
\altaffiltext{10}{Department of Astronomy and Astrophysics, 5640 South Ellis Avenue, Chicago, IL 60637}

\begin{abstract}

The first full year of operation following the commissioning year of the
Sloan Digital Sky Survey has revealed a wide variety of newly discovered
cataclysmic variables.  We show the SDSS spectra of forty-two cataclysmic
variables observed in 2002, of which thirty-five are new
classifications, four are
known dwarf novae (CT Hya, RZ Leo, T Leo and BZ UMa), one is a known CV
identified from a previous quasar survey (Aqr1) and two are known ROSAT or
FIRST discovered CVs (RX J09445+0357, FIRST J102347.6+003841). The SDSS
positions, colors and spectra of all forty-two systems are presented. In
addition, the results of follow-up studies of several of these objects
identify the orbital periods, velocity curves and polarization that
provide the system geometry and accretion properties. While most of the
SDSS discovered systems are faint ($>$ 18th mag) with low accretion rates
(as implied from their spectral characteristics), there are also a few
bright objects which may have escaped previous surveys due to changes in
the mass transfer rate.

\end{abstract}


\section{Introduction}

The commissioning year of the Sloan Digital Sky Survey (SDSS; York et al. 2000), which was
released as the Early Data Release (EDR; 
Stoughton et al. 2002),
showed the potential for scientific discoveries across a variety of
disciplines. For the field of cataclysmic variables (CVs), SDSS is able to
find many new CVs, especially
those with faint magnitudes that were missed in previous surveys with brighter
limits (Szkody et al. 2002; Paper I). This provides a more accurate picture
of the true population of CVs, without the bias of a sample based on high
accretion luminosity, and enables the study of the systems which have evolved
to the shortest orbital periods and thereby have the lowest mass transfer rates
(Howell, Rappaport \& Politano 1997). The SDSS photometry provides accurate
magnitudes, colors and positions, and the spectroscopy usually allows
unambiguous identification from the strong hydrogen Balmer and helium emission
lines that are the typical signatures of mass transfer in a close binary.   
Paper I reported 19 new discoveries from the spectra obtained 
through December 31, 2000, including high inclination systems, novalikes
and those suspected of harboring magnetic white dwarfs (see Warner 1995 for
a complete description of all types of CVs). This paper provides information
on an additional 42 CVs found from SDSS spectra obtained through 2001 
December 31. Of these 42, seven are previously 
known CVs
from prior X-ray, optical, and radio surveys, 3 were previously known objects 
but not classified as CVs, and 32 are new discoveries. 
 Followup observations on a number of
these systems include photometric light
curves,
time-resolved spectroscopy and polarimetry. The resulting analysis
 allows us to provide
a determination
of the orbital period, inclination, and magnetic field for several of the CVs.

\section{Observations and Reductions}

The SDSS imaging and spectroscopic instrumentation and reductions are
explained in detail in Paper I and in the papers by 
Fukugita et al. (1996), Gunn et al. (1998),
Lupton, Gunn, \& Szalay (1999), Hogg et al. (2001), Lupton et al. (2001),
Smith et al. (2002) and Pier et al. (2003). 
As a brief summary, SDSS photometry in 5 filters ($u,g,r,i,z$) is used to
select objects by color for later spectroscopy in the range of 
3900-6200\AA\ (blue beam) and 5800-9200\AA\ (red
beam) at a resolving power of $\sim$1800. 
The spectra are wavelength and flux-calibrated and corrected for absorption
bands, then classified as stars, galaxies and quasars. The resulting spectra
can then be searched by category, redshift or spectral characteristics. 

As explained in Paper I, the selection criteria by color
($u-g<0.45, g-r<0.7, r-i>0.30$ and $i-z>0.4$) for the CV
category primarily provide spectra of WD+M binaries, some of which are 
low mass transfer CVs where 
the underlying stars dominate the light. In addition, the color selection for
quasars and very blue objects provide spectra of
the blue CVs that are disk or white dwarf dominated.
A manual search of all the spectra for emission line objects at zero redshift
resulted in the identification of the 42 CVs listed in Table 1. This list
includes the objects that will be in Data Release 1 (DR1) as well as
several others that were found on additional plates taken in 2001. 
The magnitudes
and colors are from the psf photometry and have not been corrected for
reddening. Figure 1 shows the locations of the 42 
 CVs in SDSS color-color plots
along with the stellar locus. 

During 2001, we also conducted follow-up observations to refine the nature
of the CVs that were being discovered.
Photometry was accomplished using the University of Washington 0.76m telescope
at Manastash Ridge Observatory (MRO), using a 1024$\times$1024 CCD and a Harris
V filter and also
with the U.S. Naval Observatory Flagstaff Station (NOFS) 1m R/C telescope, 
using a 2048$\times$2048 SITe/Tektronix CCD. The
magnitudes were measured using the IRAF
\footnote{{IRAF (Image
 Reduction and Analysis
Facility) is distributed by the National Optical Astronomy Observatories, which
are operated by AURA,
Inc., under cooperative agreement with the National Science Foundation.}} task
$\it{qphot}$ and a light curve was constructed from differential magnitudes
with respect to 
comparison stars on each frame. In order to obtain the highest time
resolution with the NOFS, no filter was used. The NOFS CCD has a wide-band
response that falls between Johnson V and Cousins R, but is somewhat closer
to a V response. From separate nights of calibrated all-sky photometry using
Landolt standards, the Johnson B and V magnitudes of the comparison stars were
calibrated and used to place the photometry of the SDSS sources onto the 
Johnson V magnitude system.  

Time-resolved spectroscopy was done at the 3.5m telescope at 
Apache Point Observatory (APO) with
the Double Imaging Spectrograph in high resolution mode (resolution about
3\AA) and a 1.5 arcsec slit. Blue and red spectra were obtained on all nights 
except for 2002 May 9, when only the red spectrograph was available.
The reductions were identical to those of Paper
I and details are provided in that reference. 
For the objects with sufficient time coverage to determine a radial velocity
curve, velocities were measured with either the ``e" or ``k" routines
 in the IRAF
$\it{splot}$ package (for simple or narrow line structure) or with the
double-Gaussian method (Shafter 1983). Sinusoidal fits to these 
velocities were then used to determine $\gamma$ (systemic velocity), 
K (semi-amplitude), P (orbital period) and $\phi_{0}$ (phase of crossing from
red to blue). 

Additionally, the spectropolarimeter SPOL 
was used on the 2.2m telescope at Steward Observatory (SO) to measure the
circular polarization of a few sources to determine if they were magnetic
CVs. The dates and type of followup observations are summarized in Table 2. 

\section{Results}

For convenience, throughout the rest of this paper, we will abbreviate the
names to SDSShhmm (except for the 2 sources which have identical first 4
coordinates of 1258 so we add the degrees of declination to the name to
differentiate the two). The coordinates are accurate enough to identify the
objects easily on the Digitized Sky Survey. The few objects that have  
close companions on the sky are identified as notes to the Table.
The last column of Table 1 gives a brief comment on any notable properties of
the system. 

Figure 2 shows the SDSS spectra for all
systems.
Table 3 lists the equivalent widths and fluxes of the prominent hydrogen Balmer
and helium lines, as well as the plate and fiber number of each spectrum. 
The large strength of \ion{He}{2} in some systems allows an automatic separation into
those with possible magnetic white dwarfs (see section 3.4). 
Followup observations were
targeted first for those systems with prominent \ion{He}{2} or with prominent central
absorption lines that would signify an eclipsing system. The results for
those systems with sufficient data to accomplish a detailed classification
are presented below. 

\subsection{Previously Known Systems}

Spectra of the dwarf novae CT Hya (SDSS0851), RZ Leo (SDSS1137), T Leo 
(SDSS1138) and BZ UMa (SDSS0853), which are all
 catalogued in the Downes et al.
(2001) catalog (D01), were obtained as a result of the color 
selection algorithms. The spectra of CT Hya, T Leo and BZ UMa
show the typical characteristics of quiescent dwarf novae that are
dominated by their accretion disks (i.e. prominent Balmer and \ion{He}{1}
 emission lines
and Ca emission in the near IR). RZ Leo, on the other hand, shows strongly
doubled lines and broad absorption surrounding the emission lines, which
indicates the white dwarf, rather than an accretion disk, is contributing 
much of the blue light. 
Mennickent \& Tappert (2001; MT01) have recently identified
a spectroscopic orbital period of 1.84 hrs and comment on the large V/R 
variations throughout the orbit. They measured the velocities using a
single Gaussian fit to the entire line that resulted in a noisy radial velocity
curve. As the lines are clearly not Gaussian
(Figure 3), we attempted a better solution by using the double-Gaussian
method on the line wings, using the MT01 period. With a Gaussian width of 100 
km s$^{-1}$
and a separation of 2000 km s$^{-1}$, our solution results in a larger K
amplitude (Table 4) and a smoother radial velocity curve 
(Figure 4). 

{\it SDSS0944:\/} 
The ROSAT identified CV RXJ0944.5+0357 (Jiang et al. 2000; J00) which is SDSS0944
was also in the Sloan spectroscopic coverage. However, the 
SDSS spectrum (Figure 2) 
taken in 2001
December shows that
the flux at 5500\AA\ is 2.5 times lower (one magnitude) than  in
the J00 observations (1999 January). This is consistent with the 
magnitude differences
in the two data sets (J00 give a B magnitude of 16.1 
 whereas the SDSS 
photometry is 17.0 in g). The lower flux  results in changes in the line
emission and the domination of the secondary over the accretion disk in the
red (see section 3.5). Since both the Hamburg survey and 
the USNO-A2.0 magnitudes
are consistent with the brighter fluxes, while the SDSS and APO spectra are
consistent with the fainter magnitudes, this system either has large
variations of mass transfer on long timescales or a large orbital flux 
variation. The followup APO time-resolved spectra over 2.3 hrs show a one 
magnitude change in the continuum over this interval (Figure 5), which argues
for an orbital variation. 
The radial
velocity curve curves of both H$\beta$ and H$\alpha$ (Figure 6, Table 4) 
indicate an orbital period near
3.25 hrs, but a longer data set will be needed to pin this
down precisely. It is unusual to have an object in the 3-4 hr period range that
has a low enough $\dot{M}$ to show the underlying stars, as most systems in
this period range are high mass transfer objects with prominent accretion
disks (Warner 1995). In addition, the shape and phasing of the continuum
light curve (Figure 5) is inconsistent with either an eclipse (which would
have minimum light near phase 0) or a hot spot coming into view (which would
have maximum light near phase 0.8). Clearly, further data on this system are
needed. 

{\it SDSS1023:\/}
The recently identified FIRST source J102347.6+003841 (Bond et al. 2002)
is also among the SDSS spectra of 2001 (SDSS1023). Since only
magnetic CVs have shown radio emission, this is an excellent candidate for
a Polar (a spin-orbit synchronized AM Her system with magnetic field over
10 MG) or an Intermediate Polar (IP) system with a field near 1 MG and a
spin period for the white dwarf much less than the orbital period. 
Our 10 APO spectra obtained over
1.7 hrs show the double-peaked profile mentioned by Bond and a changing
intensity of \ion{He}{2}. The velocity curve (Figure 7 and Table 4) 
indicates the
orbital period is about 3 hrs, and the high semi-amplitude for H$\beta$ is 
typical of a 
Polar. While the period is consistent with either an IP
or a Polar, the emission lines are more compatible with an IP interpretation,
as they do not show the large equivalent widths nor the narrow components that
 are typically seen in Polars,
especially the component from the irradiated secondary which usually is
strongest near phase 0.5.
Although Bond et al. (2002) could not identify the spin period of the 
white dwarf from
their photometry, it could be shorter than their time resolution. 
Further spectroscopy throughout the entire orbit as well as spectropolarimetry 
should resolve the issue.

{\it SDSS2238:\/}
The source listed as Aqr1 in
D01 and labelled as a CV based on its identification from the Berg et al. (1992)
quasar survey also was color-selected (SDSS2238). This object shows 
\ion{He}{2} in both the SDSS and Berg et al. (1992)
 spectra and could be a magnetic
CV candidate.
APO spectra show an orbital period of 2.0 hr from the velocities (Figure 8 and 
Table 4) but 
the 3.75 hrs of MRO photometry (Table 2) shows only random
variability at the 0.2 mag level, with no large periodic variability as is
typical for Polars. Polarimetry will be needed to check the magnetic nature.

There are also 3 sources that appear in past surveys but which were not
previously identified as CVs.

{\it SDSS0131:\/}
SDSS0131 was identified as a blue object (PHL3388,
PB8928) in the survey of Berger \& Fringant (1984;BF84). Its SDSS spectrum shows strong emission lines surrounded by broad
absorption so that it is likely a system with a low disk contribution (see
section 3.5 below). The 2 hrs of followup spectra with APO indicate an
orbital period of 1.6 hrs (Table 4; Figure 9), which is consistent with a
low mass transfer rate.

{\it SDSS0802:\/}
The coordinates of SDSS0802 match those of KUV 07589+4019, which
was identified as an sdB star by Wegner \& Boley (1993). However, the
spectrum previously obtained lacked the S/N and resolution to detect the
Balmer and HeI emission lines evident in the SDSS spectrum (Figure 2). Thus,
this object is likely to be a low inclination, high mass transfer rate 
 novalike system with a bright
accretion disk. 

{\it SDSS2258:\/}
SDSS2258 was also previously identified as a blue star (PB 7412) in the BF84
survey and then classified as an emission line object
in the HK survey of Beers et al. (1996). Its coordinates are close 
(within an arcmin) to the
X-ray source RX J225834.4-094945, although this does not guarantee they are
the same object.  
The
SDSS spectrum shows a strong, narrow emission line source that is typical of
a low inclination CV. 

\subsection{High Inclination Systems}

The systems with high inclination usually show very prominent central
absorption in the Balmer lines with increasing absorption up the Balmer series.
The CVs in Figure 2 which show this effect are SDSS0407, SDSS0901, SDSS1137,
SDSS1238, SDSS1250 and SDSS1327. Photometry of SDSS1137 (RZ Leo) by Howell \&
Szkody (1988) and MT01 has shown a large orbital modulation but no eclipses.
Followup photometry on SDSS0407, SDSS0920 and 
SDSS1327 reveals eclipses in these 3 systems while time-resolved spectroscopy
provides the radial velocities as well as the periods
 (Tables
2 and 4). The most extensive observations exist for SDSS1327 (which are
described in a separate paper; Wolfe et al. 2003) and SDSS0407.

{\it SDSS0407:\/} MRO photometry first 
revealed a deeply eclipsing (2 mag) system, 
but clouds prevented the
acquisition of further data.  NOFS observations later obtained 2 eclipses
(Figure 10) that revealed an orbital period of 3.96 hrs and a strong orbital
hump modulation due to a prominent hot spot. APO spectra were
obtained for 3.5 hrs, covering an eclipse. Figure 11 shows the typical 
double-peaked profile at phase 0.5, the strong blue peak during the hump
phases and the remaining Balmer emission during the eclipse itself.
 Using the photometric eclipse period,
the double-Gauusian fitting applied to the Balmer lines gave a good fit
for H$\beta$, but large errors for H$\alpha$. 
The resulting parameters from the sine fit
to the velocities are given in
Table 4 and the velocity curve for H$\beta$ is shown in Figure 12.  

{\it SDSS0920:\/} NOFS photometry revealed eclipses for this system
even though the lines do not show the usual deep doubling. Figure 13 shows
the 2 mag deep eclipses recurring on an orbital period of 3.6 hrs. 
Unlike SDSS0407, 
there is no strong modulation hump
from a hot spot. Since the APO spectra cover less than half an orbit, a
reliable radial velocity solution could not be determined.

{\it SDSS1238:\/} The double-Gaussian fitting to the complex line profiles
of this source enabled a period determination of 1.27 hrs from both the
H$\alpha$ and H$\beta$ lines (Figure 14). The very short period and the
appearance of the spectra are very similar low mass transfer systems
 such as WZ Sge
and RZ Leo.
 
{\it SDSS1250:\/} The 2.3 hrs of APO time-resolved spectra indicate an
orbital period near 5.6 hr 
but further data are needed to
 confirm
this result. There was no noticeable eclipse within our spectral coverage,
but  
the long (15 min) 
integration times and incomplete coverage of the orbit do not allow us 
 to eliminate their 
existence. Photometry 
with better time resolution is needed to search for possible eclipses.

\subsection{Dwarf Novae}

As mentioned in Paper I, there are 3 epochs of observation for all SDSS 
spectral sources
to determine if there are large magnitude changes related to a dwarf nova
outburst or to changes from low to high mass transfer states in novalikes.
These include the SDSS photometric epoch, the SDSS spectral epoch and the
Digitized Sky Survey (DSS) epoch. A dwarf nova outburst can be distinguished from the
high state of a novalike by the spectral appearance, as a dwarf nova at
outburst shows broad absorption lines (from the thick disk), while a novalike
at a high state usually shows an enhancement of all emission, especially the
high excitation lines of \ion{He}{2} (Warner 1995).

{\it SDSS0310:\/}
Of the objects in Table 1, we can identify SDSS0310 as a bona-fide dwarf
nova, as the SDSS photometry and the APO followup spectra caught it at 
outburst (Figure 15), while
the SDSS spectrum (Figure 2) was at quiescence. Photometry by A. Henden (private
communication) has 
determined a quiescent magnitude of V=20.7. Unfortunately, followup 
APO spectra at
quiescence were too noisy to allow a good radial velocity solution and we
could not obtain a good solution from the broad absorption lines at outburst 
either, so we have no estimate of the orbital period.

Three additional systems (SDSS1146, SDSS1636 and SDSS2234) show typical
spectra of dwarf novae, although photometry/spectroscopy during an outburst
is needed to confirm the spacing and amplitude of outbursts.

{\it SDSS1146:\/} The fit to the velocities derived from APO spectra
(Table 4) reveal an orbital period near 1.6 hrs 
(Figure 16). The short orbital period and large H$\beta$ equivalent width are 
indicative of a low mass transfer
system (Patterson 1984).

{\it SDSS1636:\/}
The lack of any orbital variation during 4 hours of photometry
on SDSS1636 (Table 2) indicates this is likely a low inclination 
system. Time-resolved spectra should be able to reveal the orbital period.

{\it SDSS2234:\/} The APO spectra reveal an orbital period of 2.0 hrs
(Table 4 and Figure 17) for this 
system, which places it
right at the lower edge of the period gap.  The strong blue continuum and
emission lines are typical of accretion disk systems. 
   
Two other systems show evidence of low states (SDSS2050 was observed to be
much fainter than its SDSS photometry during MRO observations in July 2002
and SDSS2101 is fainter on the DSS than during SDSS observations). The presence
 of low states together with strong \ion{He}{2} emission makes both of these 
likely candidates for magnetic CVs (see next section). However, sufficient
followup data are not yet available to confirm the nature of these sources.
 
\subsection{Nova-likes with Strong \ion{He}{2}}

The presence of strong \ion{He}{2} emission lines often signifies a 
magnetic white dwarf, as they arise from the ionization of the accretion 
column(s) by the EUV continuum of a radial accretion shock.  If an object 
shows a large circular polarization and strong modulation of the lines and 
continuum at the orbital period, it can be be classified as a Polar. 
If it shows little or no polarization, generally 
single-peaked lines, and periodicity at plausible spin timescales of the white 
dwarf (minutes), it is considered an IP or SW Sex star 
(some SW Sex stars have recently been proposed as IPs; Rodriguez-Gil et al. 
2000).

Figure 2 and Table 3 contain 9 systems that clearly have stronger \ion{He}{2} 
lines than are expected for a non-magnetic disk system.  These, which we 
identify as the best candidates for magnetic systems, include SDSS0752, 0809, 
1023, 1327, 1446, 1553, 1700, 2050, and 2101.  SDSS0932 and 2238 may also 
qualify.  Of this group, SDSS1553 has been identified as an extremely low 
mass-accretion rate Polar (Szkody et al. 2003), and as mentioned above, Bond 
et al. (2002) have identified SDSS1023 as a FIRST radio source, with high 
probability of being a magnetic IP system.  
SDSS1327 is classified as a likely SW 
Sex star through the work of Wolfe et al. (2003). Our followup observations 
with APO, MRO, and spectropolarimetry at Steward Observatory have helped to 
pin down the nature of a few others.  These characteristics are summarized 
below.

{\it SDSS1700:\/} Photometry of SDSS1700 from MRO (Figure 18) reveals a
strongly modulated light curve with a period of 115 min., suggestive of a
Polar.  APO spectra over half the orbit give a radial velocity solution
(Table 4) with a relatively low K of 90 km s$^{-1}$ (Figure 19).  
Spectropolarimetry at the 2.3m Bok telescope (Figure 20) confirms the Polar
nature, with the broadband circular polarization varying between +2\% and +28\%
over the 2 hr period of the observations.  The constant sign of circular
polarization indicates that we see only one accretion pole through the orbit,
while the rather weakly modulated H$\alpha$ flux and modest radial velocity
amplitude suggest a rather low inclination.
In Figure 20 it is clear that the light curve variation is due to a changing
continuum shape, from rising slightly to the blue in the first spectrum, to
being strongly convex upward for the next few observations, and returning to
the initial shape by the end of the sequence.  Because these variations
correlate with the changing circular polarization, both are taken to arise from
the beaming pattern of cyclotron emission.  In the coadded polarization
spectrum (Figure 21), diffuse absorption features can be recognized near
6100\AA\ and between H$\beta$ and \ion{He}{2} $\lambda$4686.  Interpreting
these as Zeeman features of H$\alpha$ and H$\beta$, respectively, the field
strength near the accretion shock is $\sim30$ MG, so the optical spectrum is
sampling cyclotron harmonics $\sim$$5-9$.  These characteristics are rather
typical for Polars.

{\it SDSS0752:\/} Photometry obtained at NOFS coincident with our recent
XMM observation of this source revealed an orbital period of 2.7 hr (Henden,
2003, private communication). The 1.9 hrs of APO time-resolved spectra show
velocity variations on this period (Figure 22). The spectra show the
usual narrow and broad components associated with a Polar system, so it
is very likely that this object will show some polarization. Further discussion
on this source will be deferred until the analysis of the XMM results are
completed.

{\it SDSS0809:\/} The time-resolved spectra over almost 3 hrs clearly identify
an orbital period of 2.4$\pm$0.1 hr in this system (Figure 23). 
Although this is
slightly shorter than the nominal 3-4 hr range typical of SW Sex stars, the
spectra show the strong modulation and usual signature of deep absorption 
(especially in the 
\ion{He}{1} and Balmer lines near phase 0.5 (Figure 24). As there is no apparent
eclipse, this system may be one of the lower inclination SW Sex systems.  

{\it SDSS1446:\/} Spectropolarimetry at SO established a  
polarization limit of +0.38$\pm$0.27\% for SDSS1446,
indicating that this system cannot be a Polar. The APO coverage was limited
by weather to only slightly more than an hour, so further data are needed 
to establish the exact nature of this system.

{\it SDSS2050:\/} 
While the 1.8 hrs of APO spectra on 27 Sept 2002
indicate an orbital period near 2 hrs, MRO photometry on 7 Sept only shows
an overall declining brightness with some superposed variability during a
2.3 hr observation interval.
As noted in the previous section, SDSS2050 was observed to be
in a much fainter state (too faint for data acquisition) in 2002 July so
the September variability may be related to a transition state between high
and low mass transfer, or to orbital variability. Further photometric,
spectroscopic, and polarimetric data will be needed to sort out the nature of
this object. 

{\it SDSS2101:\/} While the Balmer lines are very strong in this system and
velocities could be easily measured, there was no obvious variation (within
30 km s$^{-1}$ during the 1.3 hrs of APO spectra. Further data will be needed to
determine the orbital period, but it is apparently not a short period system. 

\subsection{Systems Showing the Underlying Stars}

Six systems (SDSS0131, SDSS0137, SDSS1137, SDSS1238, SDSS2048 and SDSS2205) show
broad absorption surrounding the emission lines of H$\beta$ and higher members
of the Balmer series. Of these, two (SDSS0137 and SDSS1137) also show the
TiO bandheads of an M dwarf secondary. These are indicative of the lowest 
mass transfer systems
where the disk is so tenuous that the underlying stars provide most of the
luminosity. Two of the 6 (SDSS1137 and SDSS1238) are also high inclination
candidates from their deeply doubled lines (see above). In these cases, 
the lower disk luminosity may be
partly due to the inclination effect, since a disk seen edge on contributes
less light than when the entire disk is visible. 

Since it is also possible for a dwarf nova on the decline from outburst to
show similar absorption features,  we checked the spectral fluxes against
the SDSS photometry. There is no large difference for any of these systems
so we think the spectra are the normal quiescent states of these systems.

{\it SDSS0137:\/} Our two hours of spectral coverage of this source shows
an ultrashort orbital period of 1.4 hrs and a high amplitude velocity curve
(Figure 25 and Table 4). This is consistent with a very low mass transfer
rate system at relatively high inclination.

There are 3 other systems  which show evidence of the secondary star
from an upturn in the red portion of the spectrum and the presence of the TiO
bandhead at 7100\AA, but no evidence of the white dwarf. These are SDSS0844, 
SDSS0944 (RXJ0944.5+0357),
 and SDSS1553.
The weakness of the features in SDSS0844 and the strong blue continuum 
indicates that the disk is still dominant in this system.  SDSS1553 is the
very low accretion rate Polar mentioned above. 

The contrast of the SDSS spectrum
of SDSS0944 with that shown by Jiang et al. (2000) reveals that the Balmer
lines are less optically thick (the Balmer decrement is closer to the
recombination values) and the accretion disk has a lower density (the M star
is dominating redward of 5000\AA)
at the time of the APO observations. However, as noted above, this system
has a large amplitude orbital modulation so the appearance of the secondary
star is more noticeable at some phases than others. 

\subsection{ROSAT Correlations}

The cross-identification of the new CVs with the X-ray 
ROSAT All Sky Survey
(RASS; Voges et al. 1999, 2000) reveals that 11 are X-ray sources or have
X-ray detections close to their positions. Of these, the 3 known dwarf
novae (BZ UMa, RZ Leo, T Leo) all have X-ray emission, as well as the
ROSAT identified CV RX J094432.1+035738, and our 
newly confirmed polar SDSS1700. Table 5 lists the objects and their RASS count
rates. Since SDSS1446 has no polarization, yet is an X-ray source with
strong \ion{He}{2} emission, it may be an IP. 

\section{Conclusions}

The 42 objects here combined with the 22 in Paper I give a total of 64
CVs from 325 plates or an areal density of about 0.03 deg$^{-2}$. This 
represents a lower limit
to the density of CVs, as not all possible CV candidates (by color) are
targetted for spectra. In the future, we hope to use the plate overlap
sections to accomplish identifications by a combination of variability and 
color that will provide us with a measure of completeness.
However, even with the selection obtained, SDSS is picking up a
wide variety of new CV systems, including those with high mass transfer rates
(SW Sex stars) as well as the faintest systems with the lowest mass
transfer rates and coolest white dwarfs yet found among CVs (Szkody et al.
2003). Of the 35 new discoveries reported here, about a third show
characteristics that indicate they are likely candidates for having a
magnetic white dwarf. Another third indicate very low mass transfer rates
(from visibility of the underlying stars or from very large emission line
fluxes and low continuum). From the followup data available for 26 of the 
systems, three have shown eclipses.
The majority of the orbital periods determined so far are under the period
gap, indicating that SDSS is predominantly finding the lowest mass
transfer, shortest orbital period systems that brighter surveys have
missed e.g. the PG survey (Green et al. 1982)  and the Hamburg survey
(Hagen et al. 1995). With an aperture of 3.5m, we are currently limited in followup work
to systems brighter than 19th magnitude. It will take larger telescopes to
determine the nature of the faintest CVs being found. However, by the end
of the SDSS survey, we should have a clearer picture of the range of parameters
 and
the distribution of types of CVs in the galaxy, including both high and 
low accretion rate
systems. 

\acknowledgments

We gratefully acknowledge Michael Strauss and Patrick Hall for pointing out
some of the CV candidates and Don Schneider for useful comments on the
manuscript. 
Funding for the creation and distribution of the SDSS Archive has been provided 
by the Alfred P. Sloan Foundation, the Participating Institutions,
the National Aeronautics and Space Administration, the National Science 
Foundation, the U.S. Department of Energy, the Japanese
Monbukagakusho, and the Max Planck Society. The SDSS Web site is 
http://www.sdss.org/. Studies of magnetic stars and stellar systems at Steward
Observatory is supported by the NSF through AST 97-30792. 
The SDSS is managed by the Astrophysical Research Consortium (ARC) for the 
Participating Institutions. The Participating Institutions are The
University of Chicago, Fermilab, the Institute for Advanced Study, the 
Japan Participation Group, The Johns Hopkins University, Los Alamos
National Laboratory, the Max-Planck-Institute for Astronomy (MPIA), the 
Max-Planck-Institute for Astrophysics (MPA), New Mexico State
University, University of Pittsburgh,
 Princeton University, the United States Naval Observatory, and the 
University of Washington. 
PS and SLH also acknowledge support from NSF grant AST-0205875 and an RRF 
grant from the
UW.

\scriptsize
 \begin{deluxetable}{lccrrrrl}
\tabletypesize{\scriptsize} 
\tablewidth{0pt}
 \tablecaption{Summary of CVs with SDSS Spectra in 2001}
 \tablehead{
 \colhead{SDSS\tablenotemark{a}} &  \colhead{Date\tablenotemark{b}} & 
 \colhead{$g$} & \colhead{$u-g$} & \colhead{$g-r$} & 
 \colhead{$r-i$} &
 \colhead{$i-z$} & \colhead{Comments\tablenotemark{c}} }
 \startdata
 J013132.39$-$090122.3* & Sep 26 & 18.27 & $-$0.07 & $-$0.14 & $-$0.19 & 0.16 & PB 8928 \nl
 J013701.06$-$091234.9* & Aug 26 & 18.69 & 0.27 & 0.23 & 0.41 & 0.29 & \nl
 J031051.66$-$075500.3* & Jan 15 & 15.49 & 0.25 & $-$0.25 & $-$0.15 & $-$0.22 & DN \nl
 J040714.78$-$064425.1* & Jan 01 & 17.75 & 0.26 & 0.32 & 0.36 & 0.28 & ec \nl
 J073817.75$+$285519.7 & Nov 19  & 19.38 & 0.45 & 0.69 & 0.44 & 0.29 &  \nl
 J075240.45$+$362823.2* & Apr 18 & 17.68 & 0.22 & 0.16 & $-$0.04 & 0.03 & HeII \nl
 J080215.39$+$401047.2\tablenotemark{d} & Oct 19  & 16.69 & 0.09 & $-$0.06 & 0.01 & $-$0.02 & NL KUV07589$+$4019 \nl
 J080908.39$+$381406.2 & Dec 08  & 15.61 & 0.14 & $-$0.03 & $-$0.15 & $-$0.11 & HeII, SW Sex type \nl
 J084400.10$+$023919.3 & Nov 11  & 18.34 & $-$0.13 & 0.39 & 0.38 & 0.26 & \nl
 J085107.39$+$030834.4 & Nov 12  & 18.80 & $-$0.11 & $-$0.07 & 0.04 & 0.23 & CT Hya \nl
 J085344.00$+$574841.0* & Jan 15 & 16.39 & $-$0.46 & 0.34 & $-$0.04 & 0.22 & BZ UMa \nl
 J090103.93$+$480911.1 & Nov 25  & 19.26 & $-$0.07 & 0.11 & 0.14 & 0.14 & high i \nl
 J092009.54$+$004244.9* & Jan 20 & 17.45 & 0.01 & 0.13 & 0.11 & 0.03 & ec \nl
 J093238.21$+$010902.5* & Feb 25 & 20.31 & $-$0.58 & 0.72 & 0.45 & 0.01 &  HeII \nl
 J094431.71$+$035805.5 & Dec 23  & 16.80 & $-$0.34 & 0.60 & 0.53 & 0.31 & RX J0944.5+0357 \nl
 J101037.05$+$024915.0* & Feb 17 & 20.76 & $-$0.38 & 0.37 & $-$0.35 & 0.30 &  \nl
 J102347.67$+$003841.2* & Feb 01 & 17.99 & 1.61 & 0.56 & 0.15 & 0.06 & FIRST J102347.6+003841 \nl
 J113722.25$+$014858.6* & Mar 21 & 18.74 & $-$0.08 & 0.15 & 0.38 & 0.59 & RZ Leo \nl
 J113826.82$+$032207.1* & Mar 21 & 14.85 & $-$0.17 & $-$0.28 & 0.08 & 0.24 & T Leo \nl
 J114628.80$+$675909.7* & Feb 15 & 18.78 & $-$0.31 & $-$0.09 & 0.20 & $-$0.02 & \nl
 J123813.73$-$033933.0* & Apr 01 & 17.82 & 0.06 & $-$0.05 & $-$0.15 & $-$0.07 & high i \nl
 J124325.92$+$025547.5* & Apr 25 & 18.30 & $-$0.24 & 0.24 & $-$0.01 & 0.08 & \nl
 J125023.85$+$665525.5* & Mar 20 & 18.70 & 0.02 & 0.02 & $-$0.05 & 0.11 & high i \nl
 J125834.74$+$640823.1* & Dec 29 & 20.55 & $-$0.55 & 0.39 & $-$0.11 & 0.16 & \nl
 J125834.77$+$663551.6* & Mar 20 & 20.20 & 0.07 & $-$0.06 & $-$0.02 & 0.36 & \nl
 J132723.39$+$652854.3* & Mar 20 & 17.77 & 0.18 & 0.12 & $-$0.05 & $-$0.09 & ec,HeII, SW Sex type \nl
 J144659.95$+$025330.3* & Apr 28 & 18.20 & $-$0.25 & 0.20 & 0.11  & 0.06 & HeII,NP \nl
 J145003.12$+$584501.9* & May 27 & 20.64 & $-$0.08 & 0.23 & $-$0.32 & $-$0.03 & \nl
 J152857.86$+$034911.7* & Apr 26 & 19.53 & $-$0.49 & 0.16 & 0.18 & 0.11 & \nl
 J155331.12$+$551614.5* & May 27 & 18.49 & 1.51 & 1.06 & 0.40 & 1.00 & Polar \nl
 J163605.01$+$465204.5 & Aug 23  & 16.72 & 0.13 & $-$0.13 & $-$0.11 & $-$0.09 &  \nl
 J170053.30$+$400357.6* & Jan 19 & 19.43 & 0.09 & 0.89 & 0.45 & 0.26 & Polar \nl
 J204448.92$-$045928.8* & Aug 24 & 16.86 & 0.41 & 0.60 & 0.27 & 0.30 & \nl
 J204817.85$-$061044.8* & Aug 24 & 19.35 & $-$0.17 & 0.10 & 0.06 & 0.29 & \nl
 J205017.84$-$053626.8* & Sep 24 & 18.13 & $-$0.27 & 0.16 & 0.24 & 0.39 & HeII,H,L \nl
 J205914.87$-$061220.5* & Sep 24 & 18.38 & $-$0.16 & $-$0.01 & 0.21 & 0.20 & \nl
 J210131.26$+$105251.5 & Oct 25 & 18.08 & $-$0.23 & 0.01 & $-$0.10 & $-$0.05 & HeII,H,L \nl
 J215411.13$-$090121.7 & Oct 20  & 19.19 & $-$0.15 & 0.01 & 0.05 & 0.04 &  \nl
 J220553.98$+$115553.7 & Nov 11  & 20.07 & 0.30 & 0.00 & $-$0.18 & 0.02 &  \nl
 J223439.93$+$004127.2 & Aug 22  & 18.10 & $-$0.19 & 0.19 & 0.13 & 0.32 & \nl
 J223843.84$+$010820.7 & Aug 24  & 18.17 & $-$0.08 & $-$0.04 & $-$0.02 & 0.03 & HeII, Aqr 1 \nl
 J225831.18$-$094931.7 & Dec 15  & 15.61 & $-$0.30 & 0.17 & $-$0.02 & 0.21 & PB 7412 \nl
 \enddata
 \tablenotetext{a}{Objects marked with an asterisk are publicly available in the SDSS DR1}
\tablenotetext{b}{UT date of spectrum in 2001}
 \tablenotetext{c}{DN is a dwarf nova, ec is eclipsing, NL is a nova-like, H,L
  shows high and low brightness states, NP is not polarized}
 \tablenotetext{d}{Object is NE star of close pair}
 \end{deluxetable}

\scriptsize
\begin{deluxetable}{lccccl}
\tablewidth{0pt}
\tablecaption{Followup Data}
\tablehead{
\colhead{SDSS} & \colhead{UT Date} & \colhead{Site} &
\colhead{Time (UT)} & \colhead{Exp (s)} & \colhead{Data Obtained} }
\startdata
0131 & 2002 Dec 29 & APO & 03:35-05:30 & 600 & 11 spectra \nl
0137 & 2003 Jan 4 & APO & 03:12-05:14 & 600/900 & 8 spectra \nl
0310 & 2001 Sep 19 & APO & 10:18-11:47 & 300 & 13 spectra at outburst \nl
0407 & 2001 Jan 16 & APO & 02:30-06:06 & 600 & 19 spectra \nl
0407 & 2001 Sep 17 & MRO & 10:30-12:37 & 600 & V filter photometry \nl
0407 & 2002 Nov 14 & NOFS & 04:47-10:18 & 180 & V photometry \nl
0407 & 2002 Dec 13 & NOFS & 03:13-07:07 & 120 & V photometry \nl 
0738 & 2001 Dec 21 & APO & 10:10-12:20 & 900 & 8 spectra \nl
0752 & 2001 Oct 21 & APO & 10:23-12:18 & 900 & 7 spectra \nl
0809 & 2002 Jan 9 & APO & 10:04-12:55 & 300 & 27 spectra \nl
0901 & 2001 Dec 21 & APO & 07:45-09:58 & 900 & 8 spectra \nl
0920 & 2001 Mar 16 & APO & 07:48-09:07 & 600 & 7 spectra \nl 
0920 & 2003 Feb 7 & NOFS & 06:18-11:49 & 120 & V photometry \nl
0920 & 2003 Feb 10 & NOFS & 05:47-11:40 & 120 & V photometry \nl
0944 & 2002 Jan 9 & APO & 07:37-09:55 & 300 & 22 spectra \nl
1023 & 2001 Dec 10 & APO & 08:57-10:50 & 600 & 10 spectra \nl
1137 & 2002 May 9 & APO & 02:58-05:27 & 600 & 14 red spectra \nl
1146 & 2003 Jan 4 & APO & 10:07-13:09 & 600/900 & 11 spectra \nl
1238 & 2001 Dec 21 & APO & 12:34-13:01 & 900/600 & 2 spectra \nl
1238 & 2002 Mar 25 & APO & 05:07-07:00 & 900 & 7 spectra \nl
1243 & 2002 May 9 & APO & 05:34-07:00 & 600 & 8 red spectra \nl
1250 & 2002 May 9 & APO & 07:24-09:45 & 900 & 10 spectra \nl
1446 & 2002 May 11 & SO & 07:22-08:12  & 3000  & spectropolarimetry \nl
1446 & 2002 Jun 15 & APO & 05:37-06:45 & 600 & 6 spectra \nl
1636 & 2002 Jul 16 & MRO & 06:33-10:23 & 600 & V filter photometry \nl
1700 & 2001 Aug 7 & MRO & 04:33-06:55 & 600 & V filter photometry \nl
1700 & 2001 Aug 9 & MRO & 04:25-07:17 & 600 & V filter photometry \nl
1700 & 2001 Aug 15 & MRO & 04:25-08:54 & 600 & V filter photometry \nl
1700 & 2001 Oct 17 & APO & 02:32-03:20 & 900/1200 & 2 spectra \nl
1700 & 2002 May 9 & APO & 09:51-11:04 & 600 & 7 red spectra \nl
1700 & 2002 May 11 & SO & 09:17-11:12 & 900 & spectropolarimetry \nl 
2050 & 2002 Sep 7 & MRO & 04:00-06:19 & 600 & V filter photometry \nl
2050 & 2002 Sep 27 & APO & 04:47-06:36 & 900 & 7 spectra \nl
2101 & 2002 Oct 3 & APO & 05:07-06:26 & 600/900 & 6 spectra \nl
2234 & 2003 Jan 4 & APO & 01:05-03:02 & 600 & 11 spectra \nl
2238 & 2003 Sep 8 & MRO & 06:44-10:29 & 600 & unfiltered photometry \nl
2238 & 2002 Dec 29 & APO & 01:44-03:29 & 600/900 & 7 spectra \nl
\enddata
\end{deluxetable}

\scriptsize
\begin{deluxetable}{llrrrrrrrrrr}
\tabletypesize{\scriptsize}
\tablewidth{0pt}
\tablecaption{SDSS Spectral Line Fluxes and Equivalent Widths\tablenotemark{a}}
\tablehead{
\colhead{SDSS} & \colhead{Plate-Fiber} & \multicolumn{2}{c}{H$\gamma$} & 
\multicolumn{2}{c}{H$\beta$} &
\multicolumn{2}{c}{H$\alpha$} &
\multicolumn{2}{c}{He4471} & \multicolumn{2}{c}{HeII4686}\\
\colhead{} & \colhead{} & \colhead{F} & \colhead{EW} & \colhead{F} &
\colhead{EW} & \colhead{F} & \colhead{EW} & \colhead{F} & \colhead{EW} &
\colhead{F} & \colhead{EW} } 
\startdata
0131 & 662-384 & 4.7 & 25 & 5.9 & 44 & 9.2 & 147 & 0.6 & 2 &  & \nl
0137 & 662-552 & 1.5 & 12 & 2.3 & 21 & 3.4 & 38 & 0.4 & 3 &  & \nl
0310 & 459-222 & 0.5 & 31 & 0.6 & 50 & 1.1 & 138 & & & & \nl
0407 & 465-226 & 8.1 & 18 & 5.8 & 12 & 10.7 & 30 & & & & \nl
0738 & 754-34 & 3.3 & 65 & 3.5 & 56 & 4.9 & 57 & 0.5 & 9 & & \nl
0752 & 543-65 & 2.8 & 24 & 2.4 & 26 & 2.0 & 35 & 0.7 & 6 & 1.4 & 14 \nl
0802 & 544-423 & 1.7 & 2 & 2.0 & 2 & 2.8 & 7 & & & & \nl
0809 & 758-570 & 24 & 6 & 30 & 9 & 35 & 24 & 3.8 & 1 & 16 & 5 \nl
0844 & 564-197 & 9.6 & 73 & 10 & 88 & 9.4 & 101 & 2.3 & 18 & 0.7 & 6 \nl
0851 & 565-359 & 7.5 & 62 & 8.7 & 87 & 9.0 & 141 & 1.5 & 12 & & \nl
0853 & 483-332 & 108 & 76 & 143 & 124 & 171 & 195 & 31 & 22 & 11 & 9 \nl
0901 & 764-589 & 3.1 & 37 & 3.8 & 58 & 5.2 & 143 & 0.3 & 4 & & \nl
0920 & 473-98 & 2.4 & 35 & 3.0 & 46 & 4.1 & 91 & 0.5 & 7 & 0.7 & 10 \nl
0932 & 475-66 & 2.4 & 42 & 2.6 & 52 & 2.7 & 72 & 0.7 & 12 & 0.5 & 9 \nl
0944 & 570-512 & 16 & 47 & 18 & 50 & 19 & 39 & 2.8 & 8 & & \nl
1010 & 502-444 & 1.9 & 104 & 2.7 & 186 & 5.8 & 553 & 0.3 & 18 & & \nl
1023 & 272-461 & 16 & 9 & 8.3 & 6 & 13 & 14 & 8.7 & 5 & 7.8 & 5 \nl
1137 & 513-70 & 2.6 & 10 & 3.5 & 20 & 8.3 & 78 & & & & \nl
1138 & 513-562 & 288 & 72 & 327 & 111 & 322 & 173 & 65 & 18 & 21 & 7 \nl
1146 & 492-454 & 7.2 & 62 & 7.7 & 80 & 8.2 & 120 & 1.5 & 13 & 0.5 & 5 \nl
1238 & 335-85 & 2.0 & 6 & 4.8 & 20 & 9.8 & 84 & 0.5 & 1.3 &  & \nl
1243 & 522-325 & 21 & 89 & 23 & 114 & 23 & 169 & 4.3 & 21 & 1.2 & 6 \nl
1250 & 495-238 & 2.3 & 18 & 3.6 & 39 & 4.6 & 85 & 0.6 & 5 & & \nl
1258+64 & 602-264 & 1.1 & 10 & 1.2 & 14 & 1.4 & 32 &  & 0.4 & 5 \nl
1258+66 & 495-148 & 5.4 & 99 & 5.1 & 111 & 4.3 & 141 & 1.2 & 24 & 0.4 & 10 \nl
1327 & 496-83 & 7.0 & 31 & 6.9 & 32 & 6.4 & 51 & 2.9 & 13 & 4.0 & 18 \nl
1446 & 537-454 & 17 & 44 & 12 & 42 & 7.7 & 46 & 3.9 & 11 & 7.0 & 23 \nl
1450 & 610-349 & 0.7 & 36 & 1.5 & 100 & 2.6 & 346 & & & & \nl
1528 & 592-601 & 4.2 & 46 & 4.2 & 60 & 4.1 & 98 & 1.0 & 12 & & \nl
1553 & 619-437 &  & & & 0.1 & 3 & & & & \nl
1636 & 627-479 & 8.0 & 7 & 8.9 & 10 & 11 & 24 & 2.4 & 2 & 1.1 & 1 \nl
1700 & 633-574 & 2.5 & 10 & 2.9 & 12 & 2.8 & 16 & 0.4 & 1.6 & 4.0 & 1 \nl
2044 & 635-387 & 17 & 36 & 17 & 30 & 23 & 37 & 3.6 & 7 & & \nl
2048 & 635-127 & 0.9 & 16 & 1.8 & 42 & 2.7 & 103 & 0.2 & 3 & & \nl
2050 & 636-331 & 17 & 95 & 18 & 109 & 11 & 83 & 3.3 & 20 & 11 & 70 \nl
2059 & 636-599 & 8.1 & 76 & 7.9 & 79 & 7.8 & 101 & 1.9 & 18 & & \nl
2101 & 727-324 & 14 & 66 & 12 & 65 & 8.2 & 90 & 3.6 & 21 & 3.7 & 23 \nl
2154 & 716-299 & 1.3 & 6 & 1.8 & 11 & 1.7 & 21 & & 0.4 & 2 \nl
2205 & 734-30 & 0.5 & 13 & 1.0 & 38 & 2.3 & 136 & & & & \nl
2234 & 376-631 & 20 & 40 & 20 & 55 & 16 & 68 & 2.4 & 5 & & \nl
2238 & 377-540 & 2.1 & 8 & 2.6 & 13 & 3.1 & 34 & 0.3 & 1.5 & 0.7 & 3 \nl
2258 & 725-306 & 122 & 42 & 137 & 52 & 100 & 61 & 35 & 13 & 12 & 5 \nl 
\enddata
\tablenotetext{a}{Fluxes are in units of 10$^{-15}$ ergs cm$^{-2}$ s$^{-1}$,
equivalent widths are in units of \AA}
\end{deluxetable}

\scriptsize
\begin{deluxetable}{lccccccl}
\tablewidth{0pt}
\tablecaption{Radial Velocity Solutions}
\tablehead{
\colhead{SDSS} & \colhead{Line} & \colhead{P (min)} & 
\colhead{$\gamma$} & \colhead{K (km/s)} &
\colhead{T$_{0}$ (JD245+)} & \colhead{$\sigma$} & 
\colhead{Method\tablenotemark{a}} }
\startdata
0131 & H$\beta$ & 98 & -63$\pm$1 & 29$\pm$5 & 2637.694 & 11 & dG (1400) \nl
0131 & H$\alpha$ & 98 & 50.1$\pm$0.3 & 25$\pm$2  & 2637.696 & 5 & dG (1400) \nl
0137 & H$\beta$ & 86 & 21$\pm$2 & 118$\pm$13 & 2643.687 & 16 & dG (1300) \nl
0137 & H$\alpha$ & 80 & 48$\pm$2 & 98$\pm$8 & 2643.673 & 14 & dG (1500) \nl
0407 & H$\beta$ & 238 & 85$\pm$2 & 122$\pm$15 & 1925.602 & 37 & dG (1700) \nl
0407 & H$\alpha$ & 238 & 77$\pm$4 & 69$\pm$29 & 1926.618 & 72 & dG (1400) \nl
0738 & H$\beta$ & 127 & 75$\pm$16 & 176$\pm$43 & 2264.916 & 51 & e \nl
0738 & H$\alpha$ & 125 & 84$\pm$12 & 264$\pm$74 & 2264.962 & 100 & e \nl
0752 & H$\beta$ & 162 & 3$\pm$8 & 78$\pm$15 & 2203.848 & 21 & dG (300) \nl
0809 & H$\beta$ & 143 & -14$\pm$1 & 128$\pm$9 & 2283.967 & 28 & dG (1000) \nl
0809 & H$\alpha$ & 143 & 109$\pm$1 & 84$\pm$14 & 2283.977 & 45 & dG (900) \nl
0944 & H$\beta$ & 195 & 11$\pm$3 & 69$\pm$5 & 2283.869 & 19 & e \nl
0944 & H$\alpha$ & 195 & 87$\pm$5 & 119$\pm$11 & 2283.850 & 37 & e \nl  
1023 & H$\beta$ & 182 & -194$\pm$14 & 190$\pm$18 & 2253.841 & 36 & e \nl
1023 & H$\alpha$ & 182 & -4$\pm$16  & 84$\pm$24 & 2253.861 & 40 & e \nl 
1137 & H$\alpha$ & 113 & -25$\pm$2 & 126$\pm$12 & 2403.677 & 30 & dG (2000) \nl
1146 & H$\beta$ & 99 & -24$\pm$2 & 120$\pm$18 & 2643.960 & 38 & dG (2100) \nl
1146 & H$\alpha$ & 95 & -2.8$\pm$0.2 & 69$\pm$14 & 2643.964 & 31 & dG (2000) \nl
1238 & H$\beta$ & 76 & -32$\pm$10 & 202$\pm$41 & 2358.708 & 68 & e \nl
1238 & H$\alpha$ & 76 & -65$\pm$2 & 93$\pm$9 & 2358.705 & 15 & e \nl
1700 & H$\alpha$ & 115 & -66$\pm$3 & 89$\pm$6 & 2403.899 & 6 & e \nl
2050 & H$\alpha$ & 138 & 86$\pm$3 & 38$\pm$12 & 2544.748 & 18 & e \nl
2234 & H$\beta$ & 120 & -19.6$\pm$0.3 & 65$\pm$4 & 2643.524 & 7 & dG (1500) \nl
2234 & H$\alpha$ & 124 & 23.5$\pm$0.3 & 108$\pm$4 & 2643.523 & 8 & dG (1900) \nl
2238 & H$\beta$ & 121 & -134$\pm$6 & 99$\pm$14 & 2637.613 & 21 & dG (1300) \nl
2238 & H$\alpha$ & 121 & -91$\pm$16 & 147$\pm$33 & 2637.625 & 29 & dG (2200) \nl
\enddata
\tablenotetext{a}{dG = double Gaussian (value in parenthesis is separation in 
km/s),
 e = centroid in splot}
\end{deluxetable}

\begin{deluxetable}{lccl}
\tablewidth{0pt}
\tablecaption{ROSAT Detections}
\tablehead{
\colhead{SDSS} & \colhead{ROSAT (c s$^{-1}$)\tablenotemark{a}} & \colhead{Exp (sec)}
& \colhead{RXS-J} }
\startdata
0853 & 0.41$\pm$0.04 & 394 & 085343.5+574846=BZ UMa \nl
0944 & 0.09$\pm$0.02 & 243 & 094432.1+035738 \nl
1137 & 0.03$\pm$0.01 & 243 & 113723.5+014847=RZ Leo \nl
1138 & 0.66$\pm$0.06 & 410 & 113826.8+032210=T Leo \nl
1146 & 0.023$\pm$0.007 & 689 & 114631.3+675932 \nl
1243 & 0.125$\pm$0.035 & 141 & 124326.5+025603 \nl
1258 & 0.03$\pm$0.01 & 477 & 125837.1+663608 \nl
1446 & 0.04$\pm$0.01 & 345 & 144700.9+025344 \nl
1700 & 0.07$\pm$0.01 & 744 & 170053.7+400354 \nl
2050 & 0.04$\pm$0.01 & 329 & 205018.4-053631 \nl
2258 & 0.17$\pm$0.04 & 145 & 225834.4-094945 \nl
\enddata
\tablenotetext{a}{For a 2 keV bremsstrahlung spectrum, 1 c s$^{-1}$ corresponds to a
0.1-2.4 keV flux of about 7$\times10^{-12}$ ergs cm$^{-2}$ s$^{-1}$}
\end{deluxetable}
\clearpage

%

\begin{figure}
\plotone{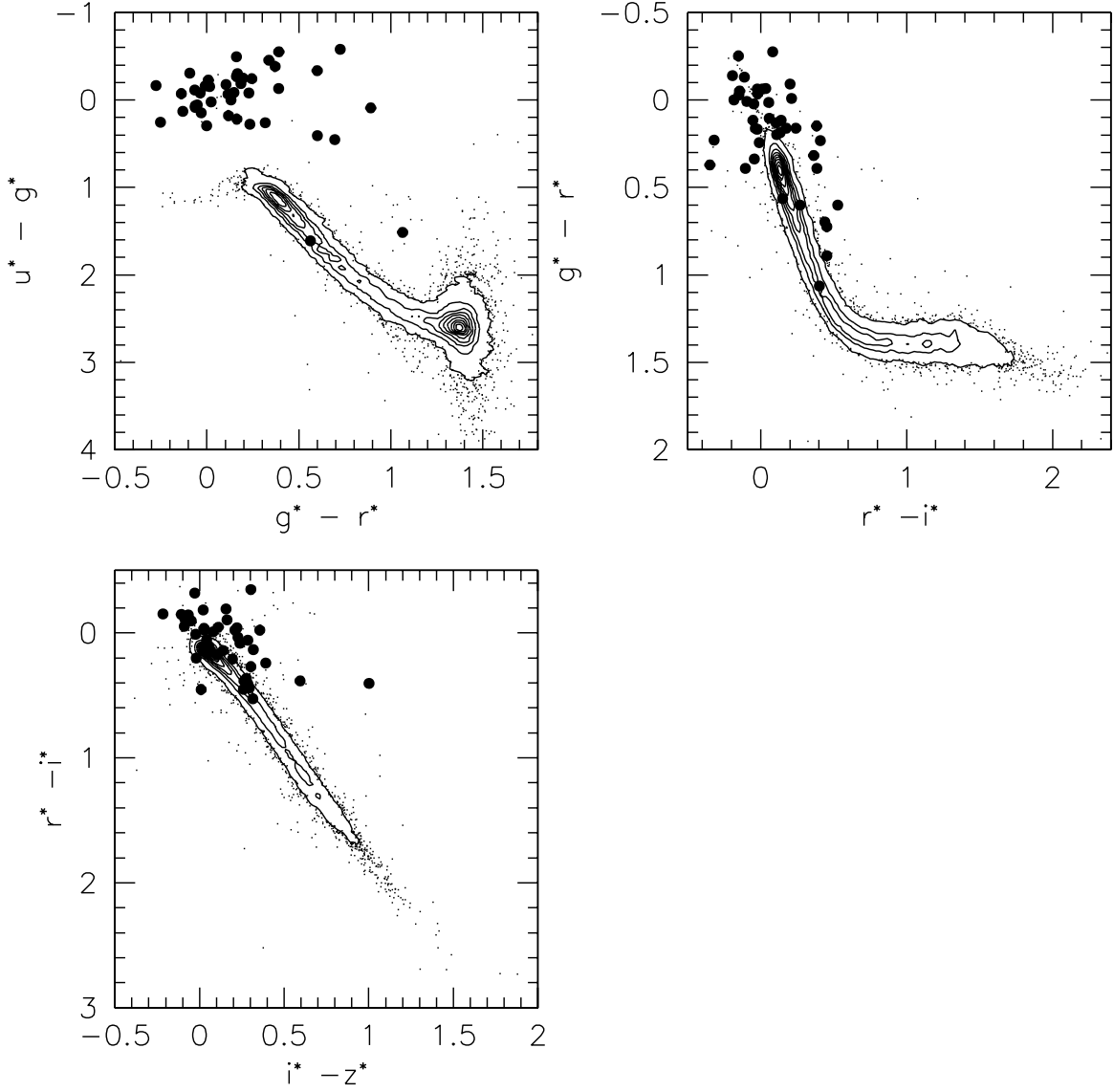}
\figcaption
{SDSS color-color plots of the objects in Table 1. Filled circles are 
the CVs, and small dots and contours (at intervals of 10\% of the peak)
 are stars defining the stellar locus.} 
\end{figure}

\begin{figure}
\resizebox*{0.9\textwidth}{!}{\includegraphics{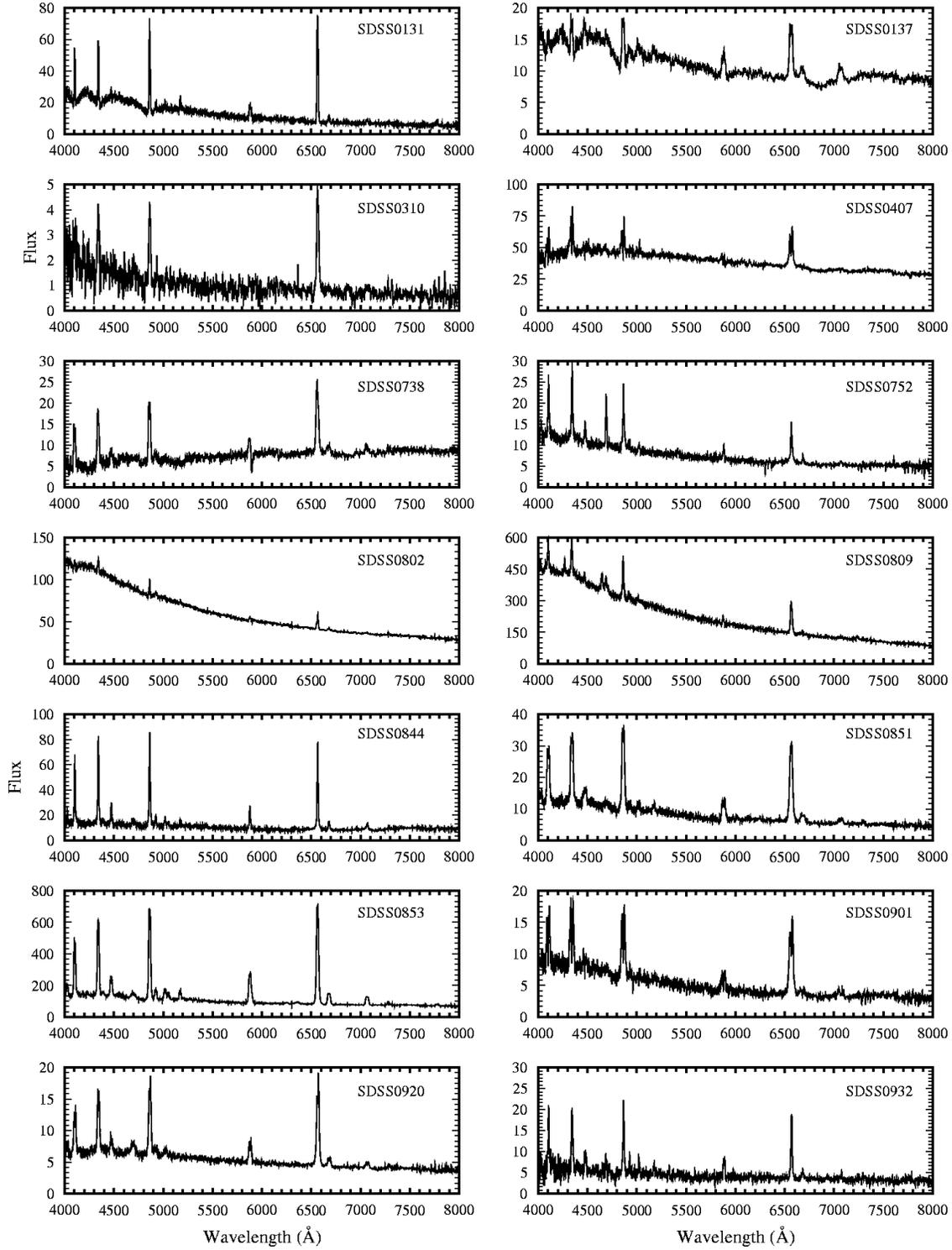}}
\figcaption{SDSS spectra of the newly discovered CVs in the interval from 4000-8000\AA. The flux scale is in units of
flux density 10$^{-17}$ ergs cm$^{-2}$ s$^{-1}$ \AA$^{-1}$.}
\end{figure}

\setcounter{figure}{1}
\begin{figure}

\figcaption{SDSS spectra of the newly discovered CVs in the interval from 4000-8000\AA\ (continued).}
\resizebox*{0.9\textwidth}{!}{\plotone{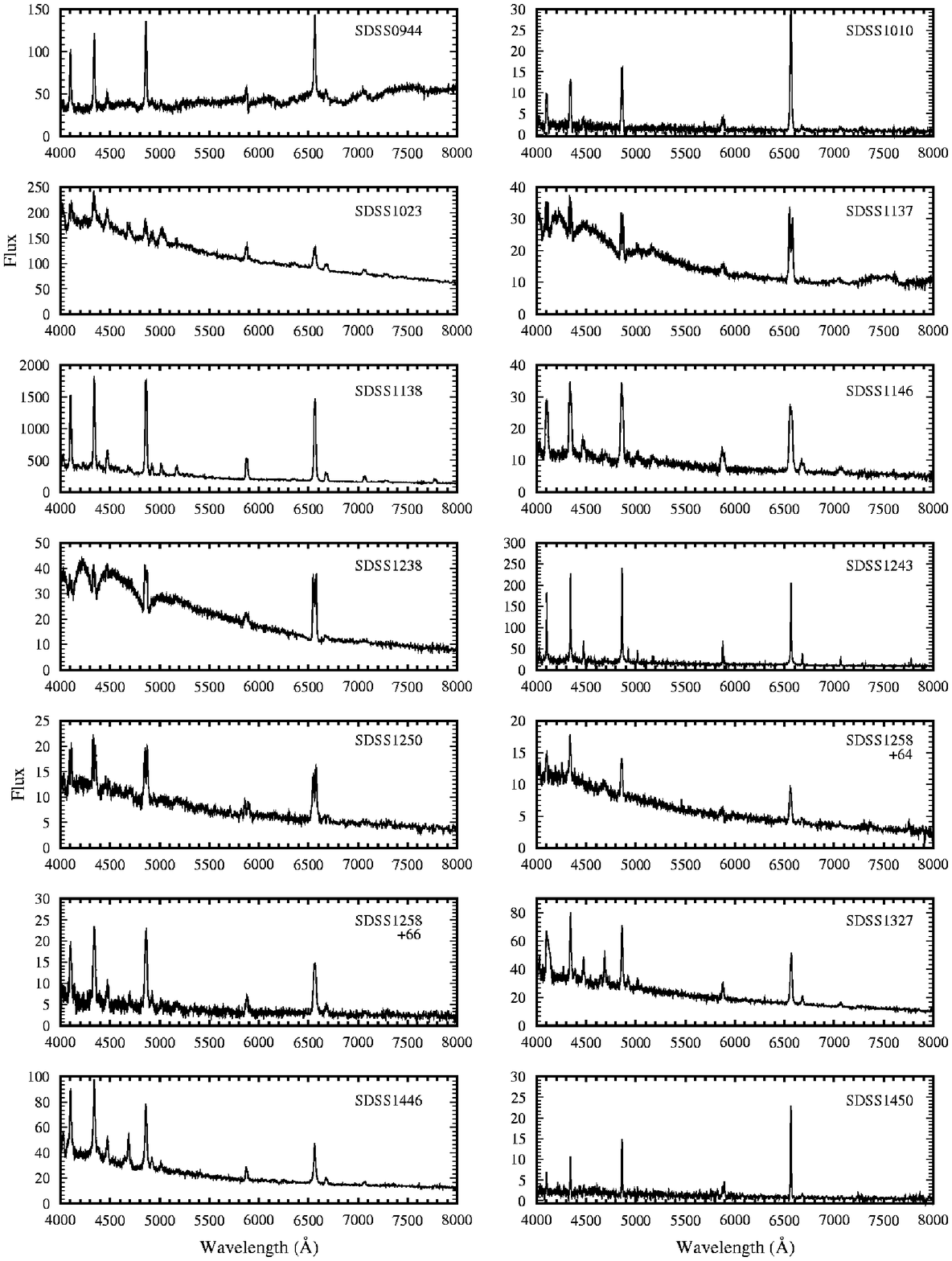}}

\end{figure}

\setcounter{figure}{1}
\begin{figure}
\figcaption{SDSS spectra of the newly discovered CVs in the interval from 4000-8000\AA\ (continued).}
\resizebox*{0.9\textwidth}{!}{\plotone{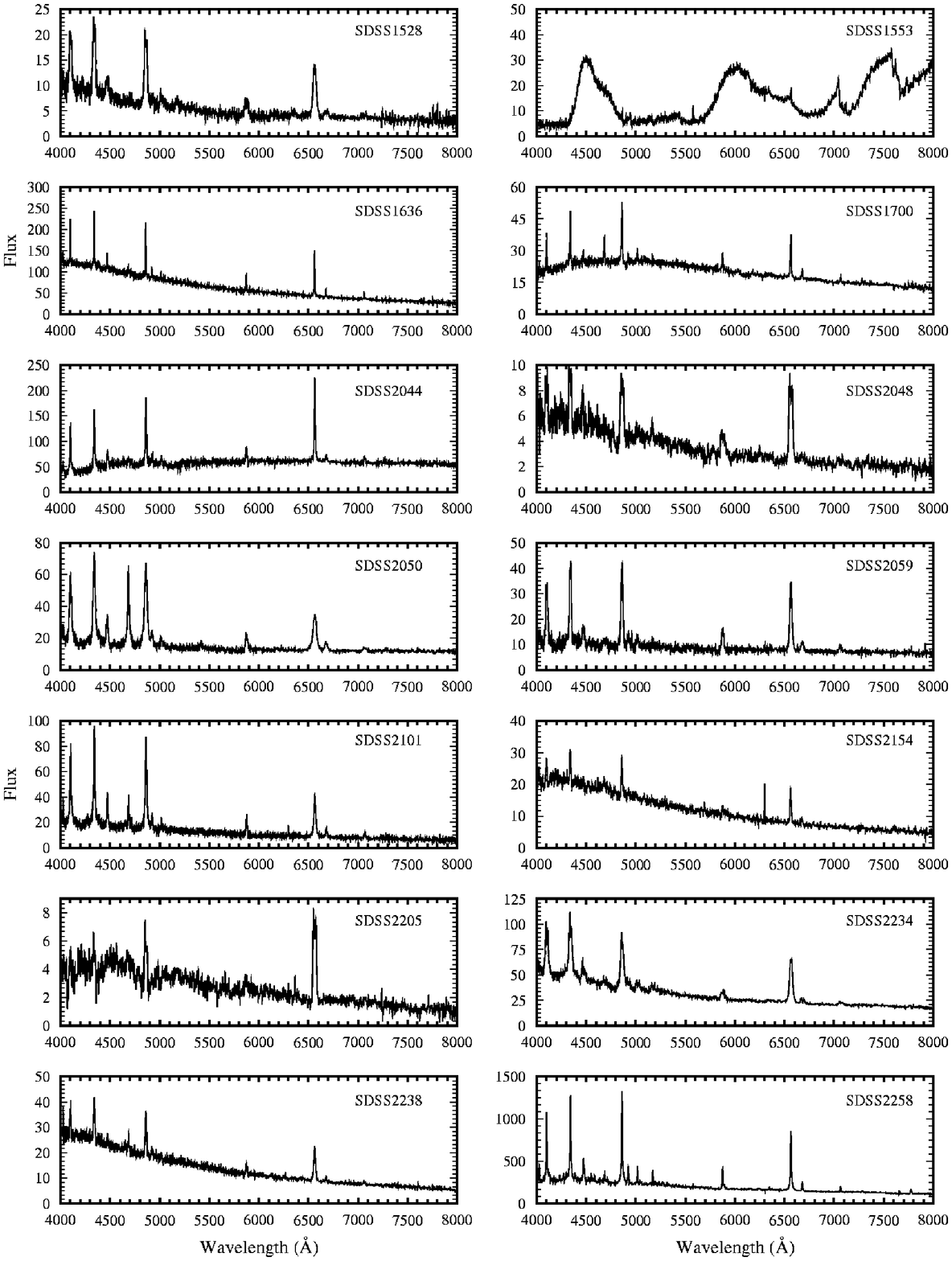}}
\end{figure}

\begin{figure}
\plotone{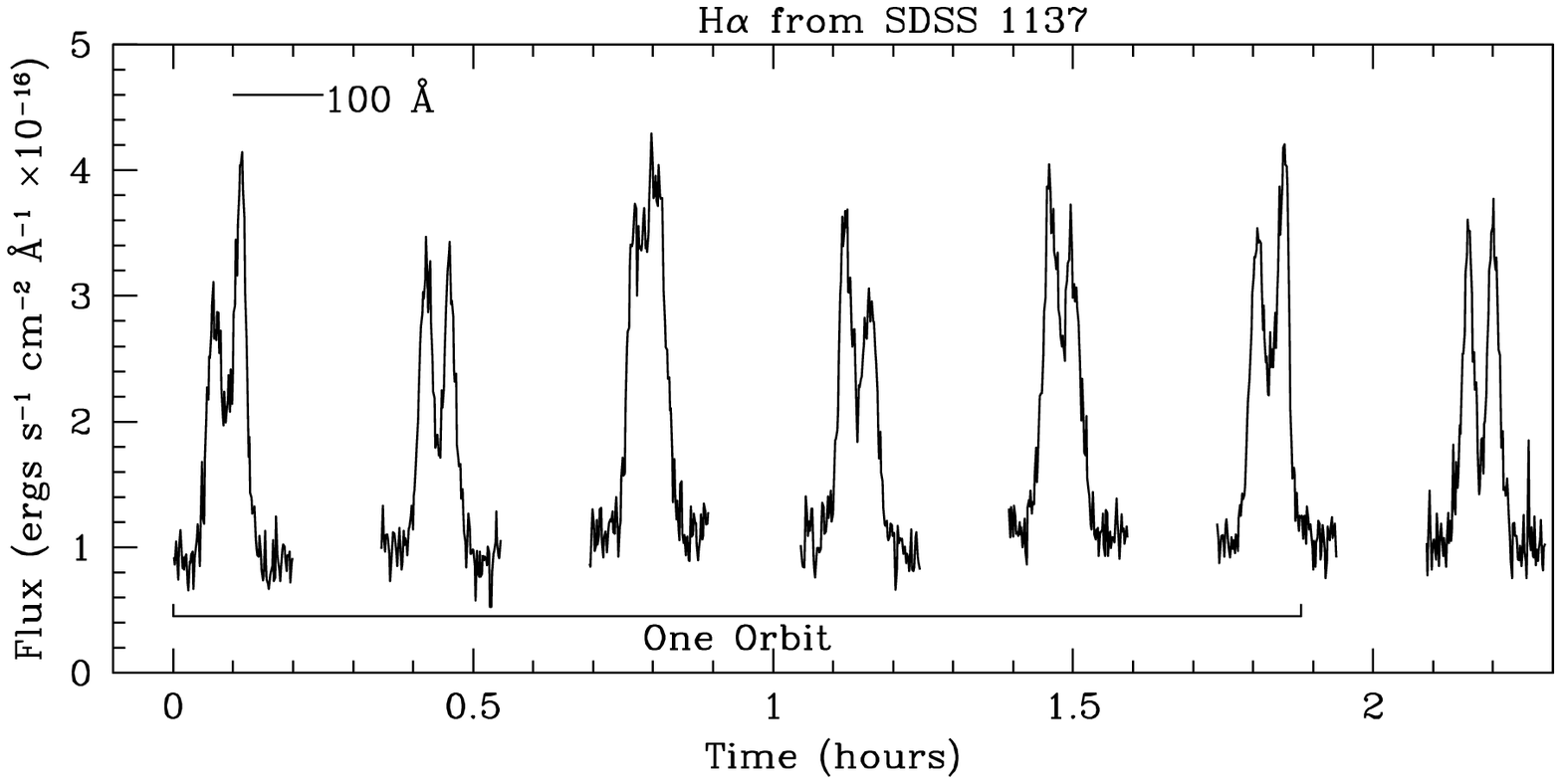}
\figcaption{Time-resolved APO spectra of RZ Leo (SDSS1137), showing the large changes
in the blue and red components of H$\alpha$ throughout the orbit.}

\end{figure}
\begin{figure}
\plotone{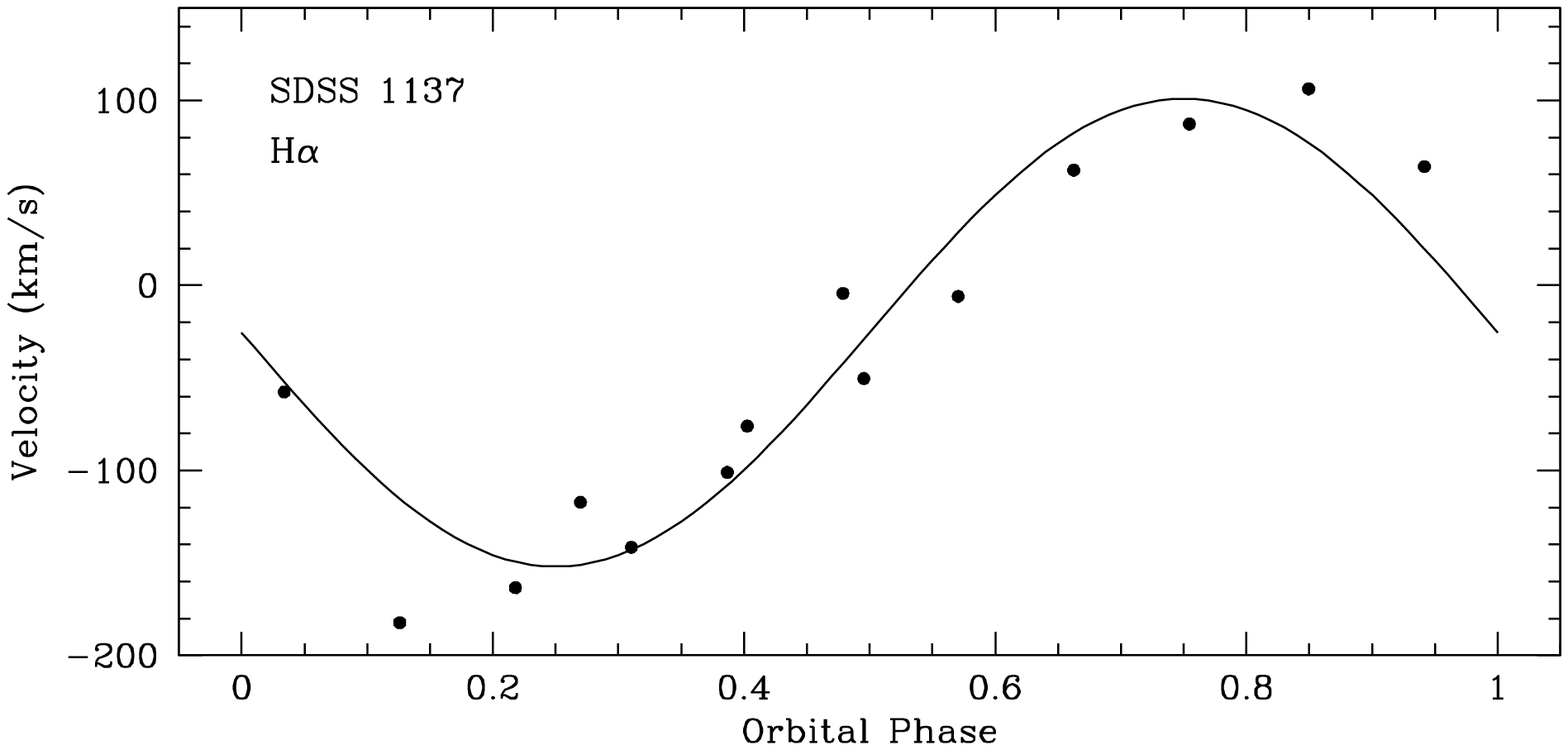}
\figcaption{Radial velocity data of RZ Leo obtained from using a double-Gaussian
fitting to the line wings, with the best-fit solution from Table 4 shown as
the solid curve.}
\end{figure}
\begin{figure}
\plotone{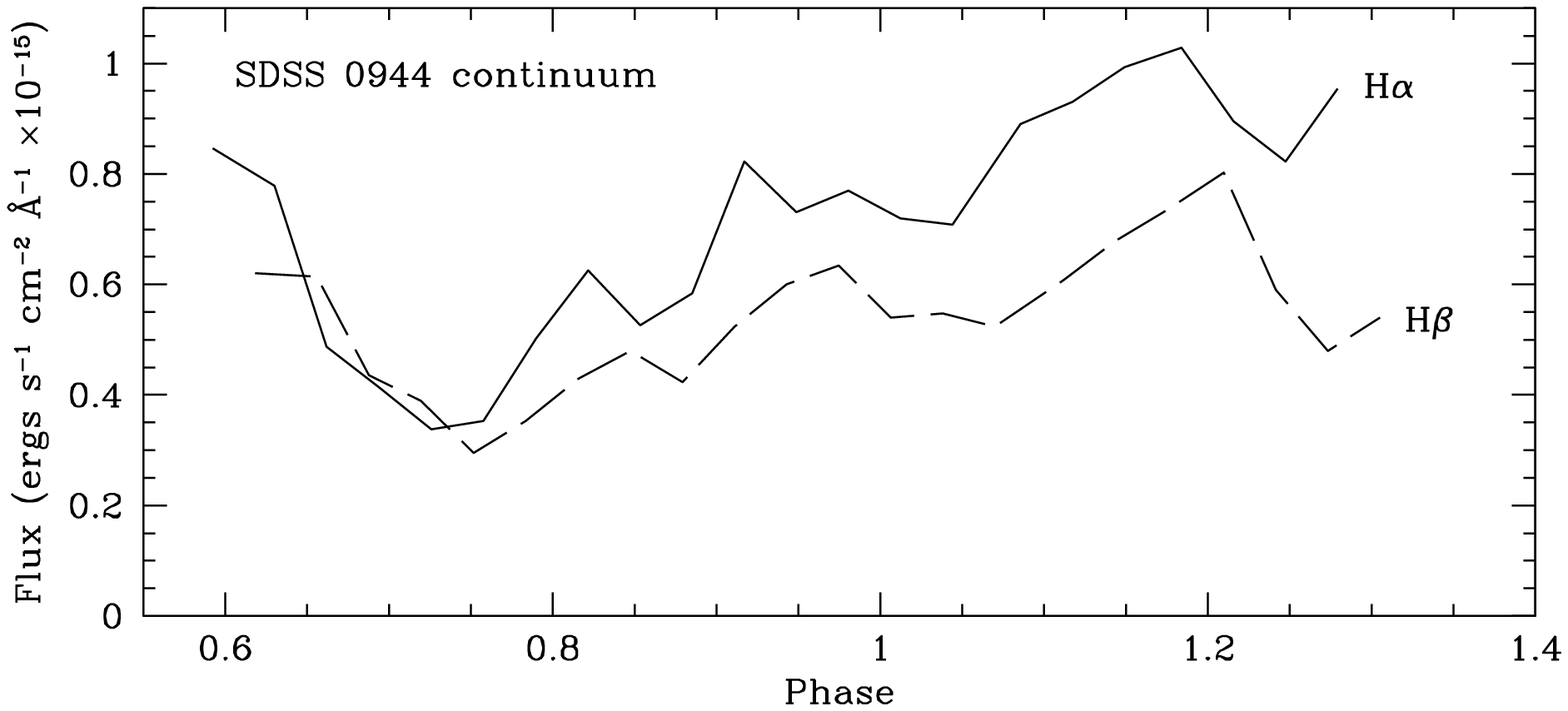}
\figcaption{Blue and red continuum changes in SDSS0944
throughout the orbit as evident from APO spectra.} 
\end{figure}
\begin{figure}
\plotone{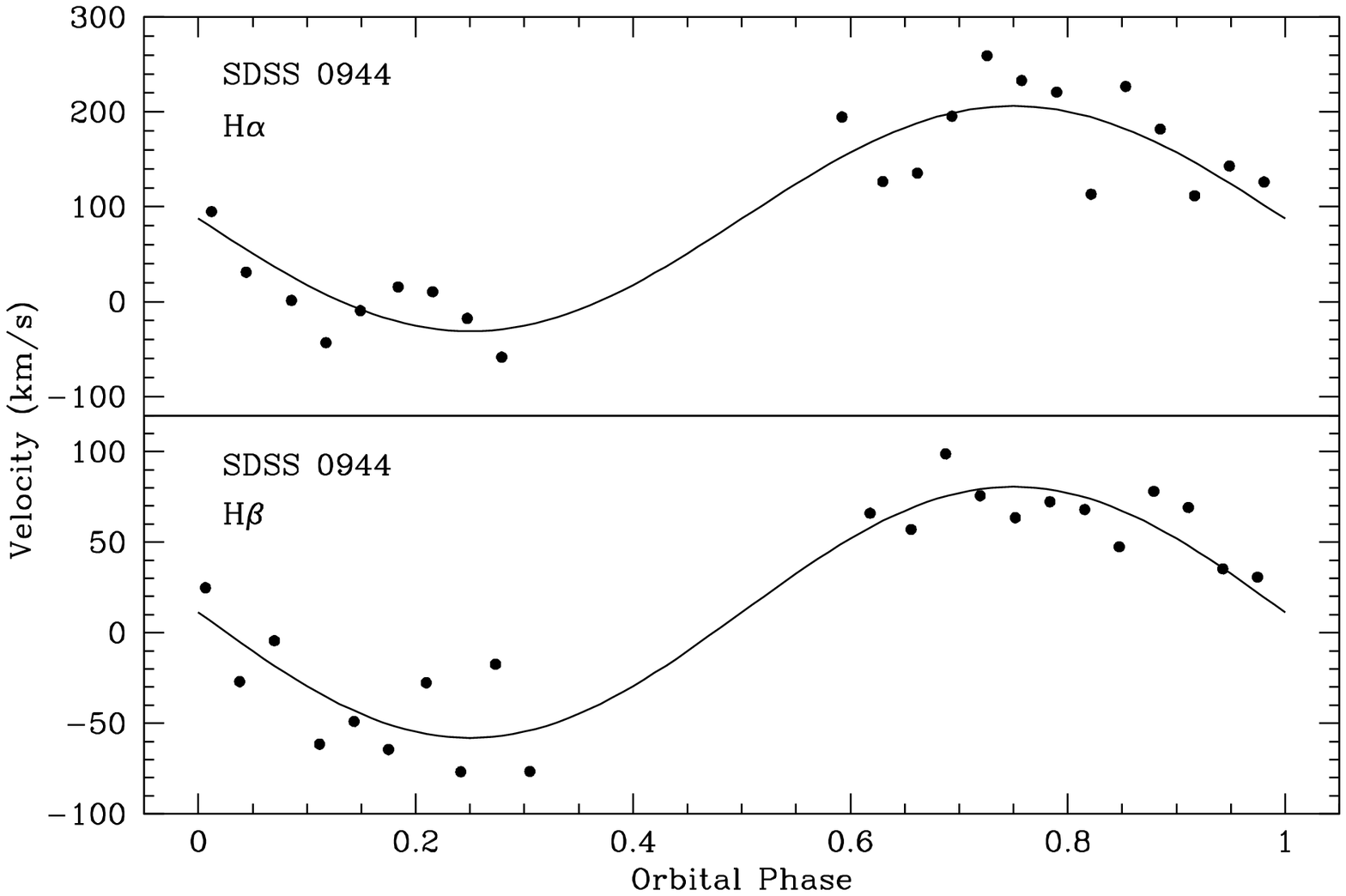}
\figcaption{Velocity curves of SDSS0944 with the best fit sinusoids (Table 4)
plotted on the data.}
\end{figure}
\begin{figure}
\plotone{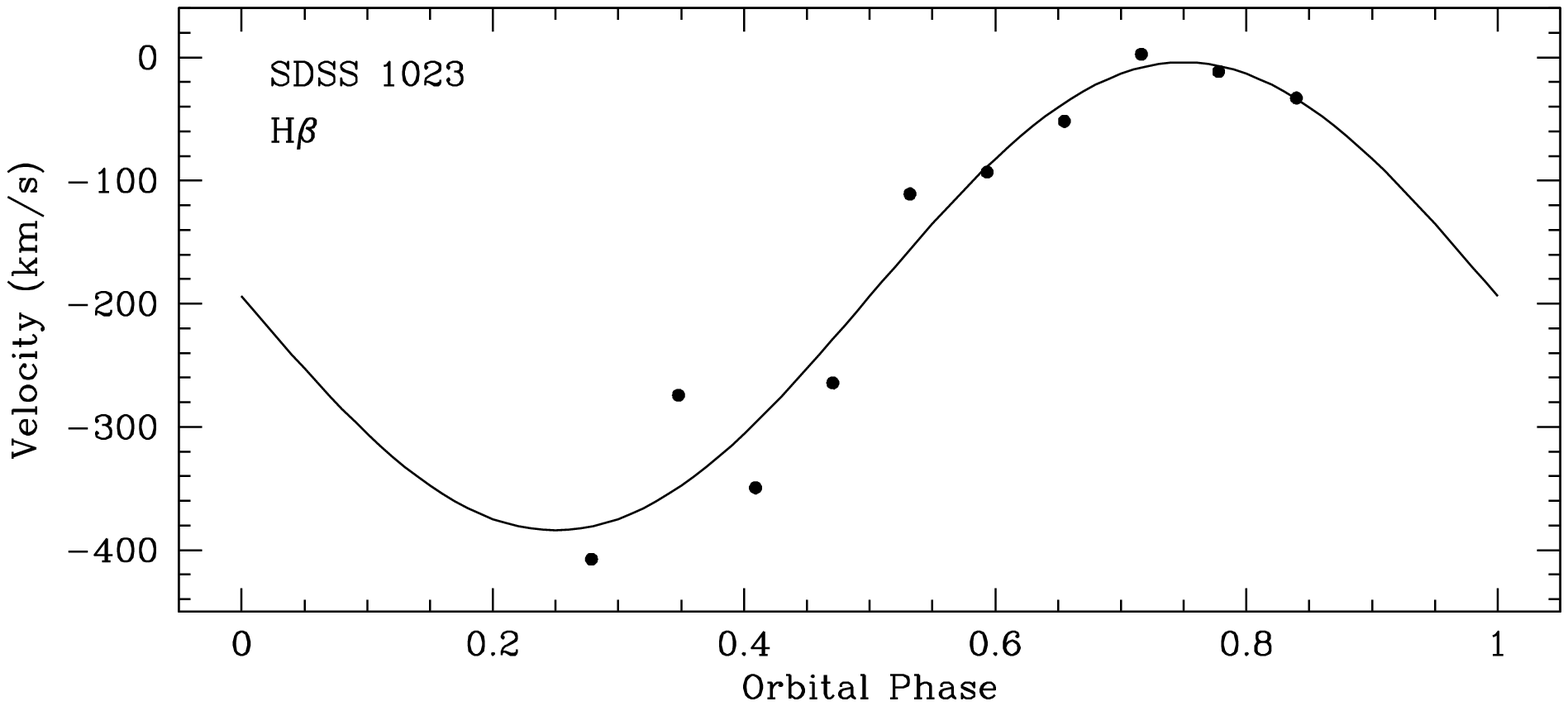}
\figcaption{Velocity curve of SDSS1023 with the best fit sinusoid superposed.}
\end{figure}

\begin{figure}
\plotone{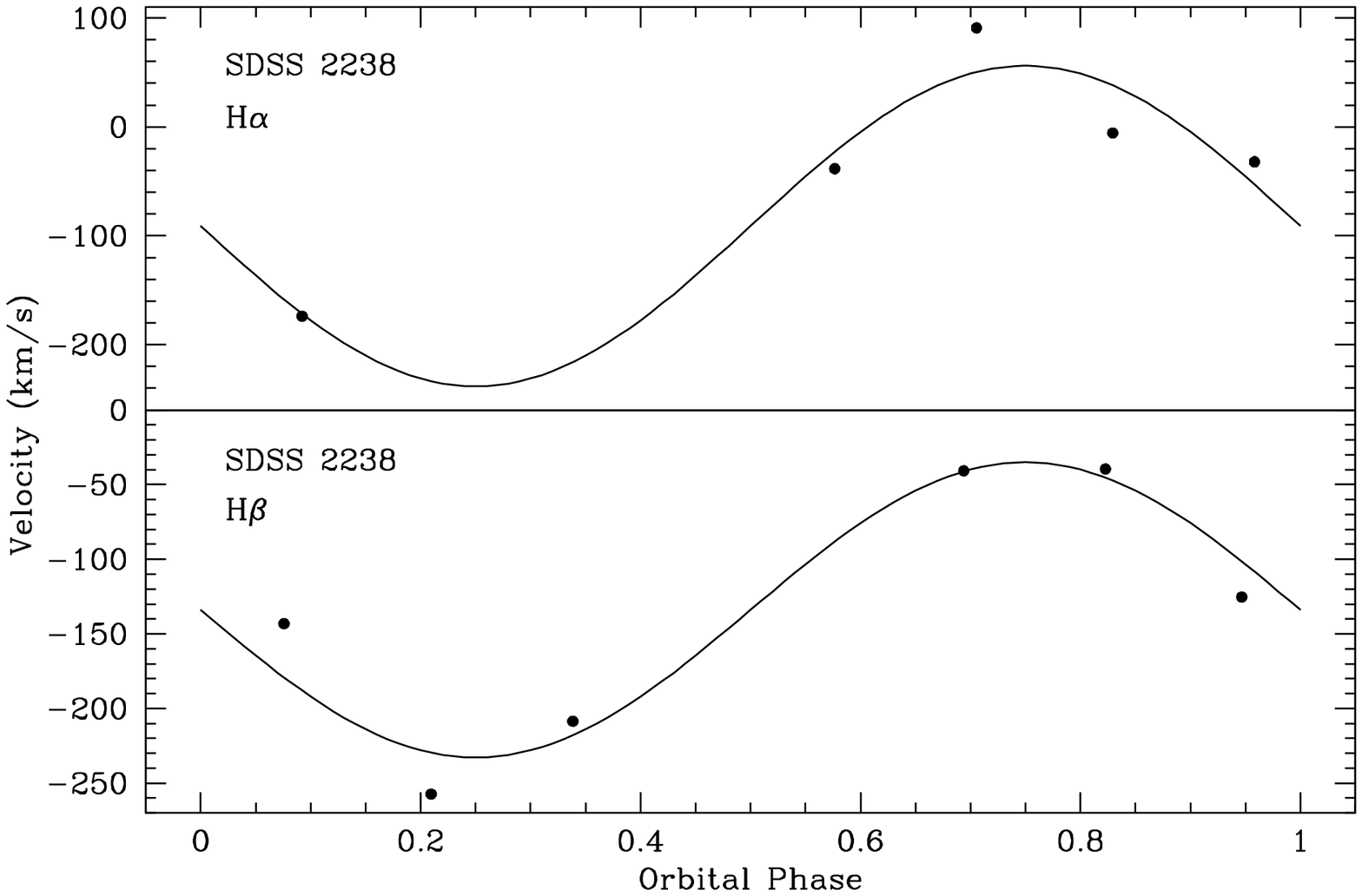}
\figcaption{Velocity curves of SDSS2238 with the best fit sinusoids superposed.}\end{figure}

\begin{figure}
\plotone{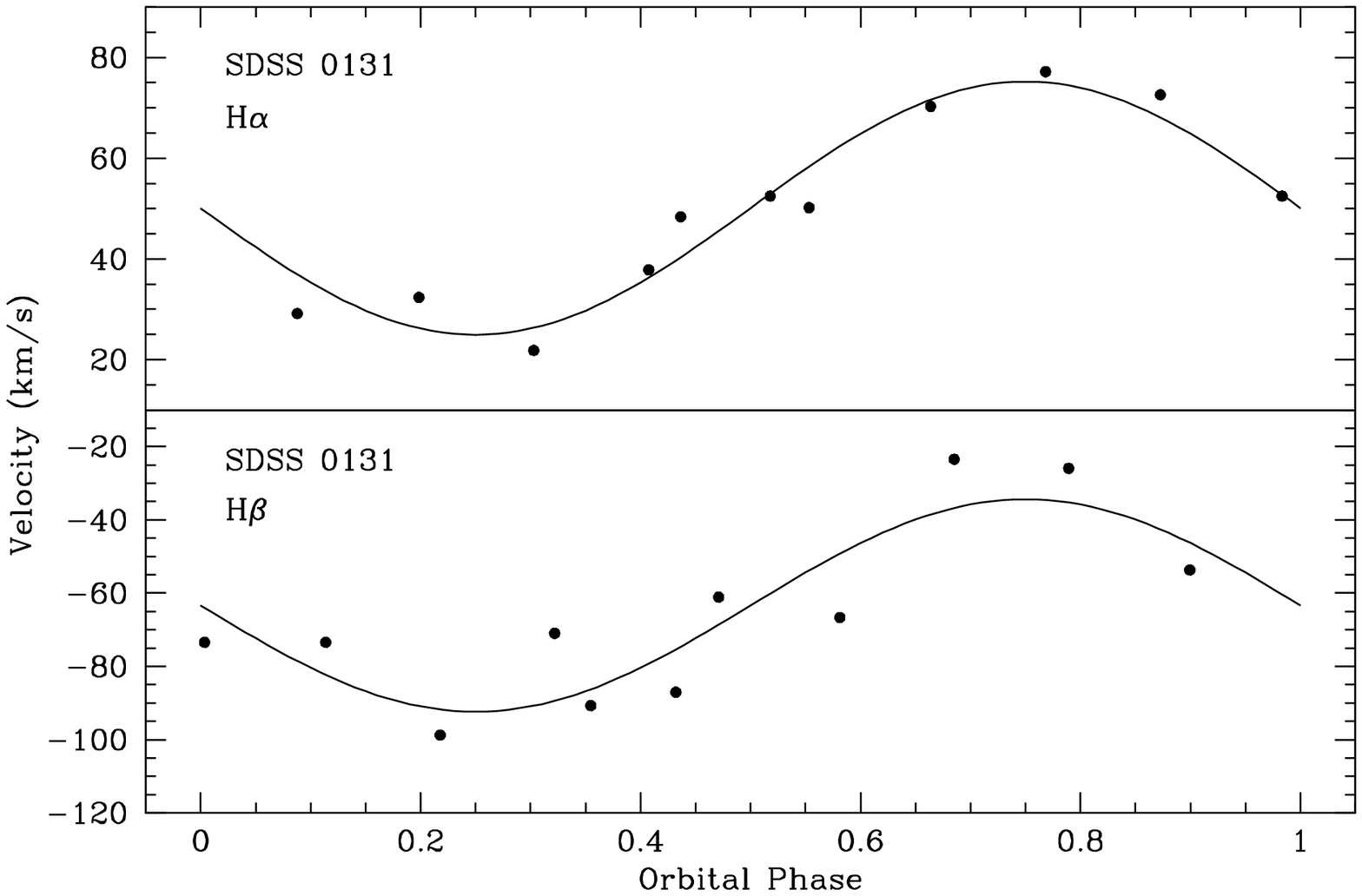}
\figcaption{Velocity curves of SDSS0131 with the best fit sinusoids superposed.}\end{figure}

\begin{figure}
\plotone{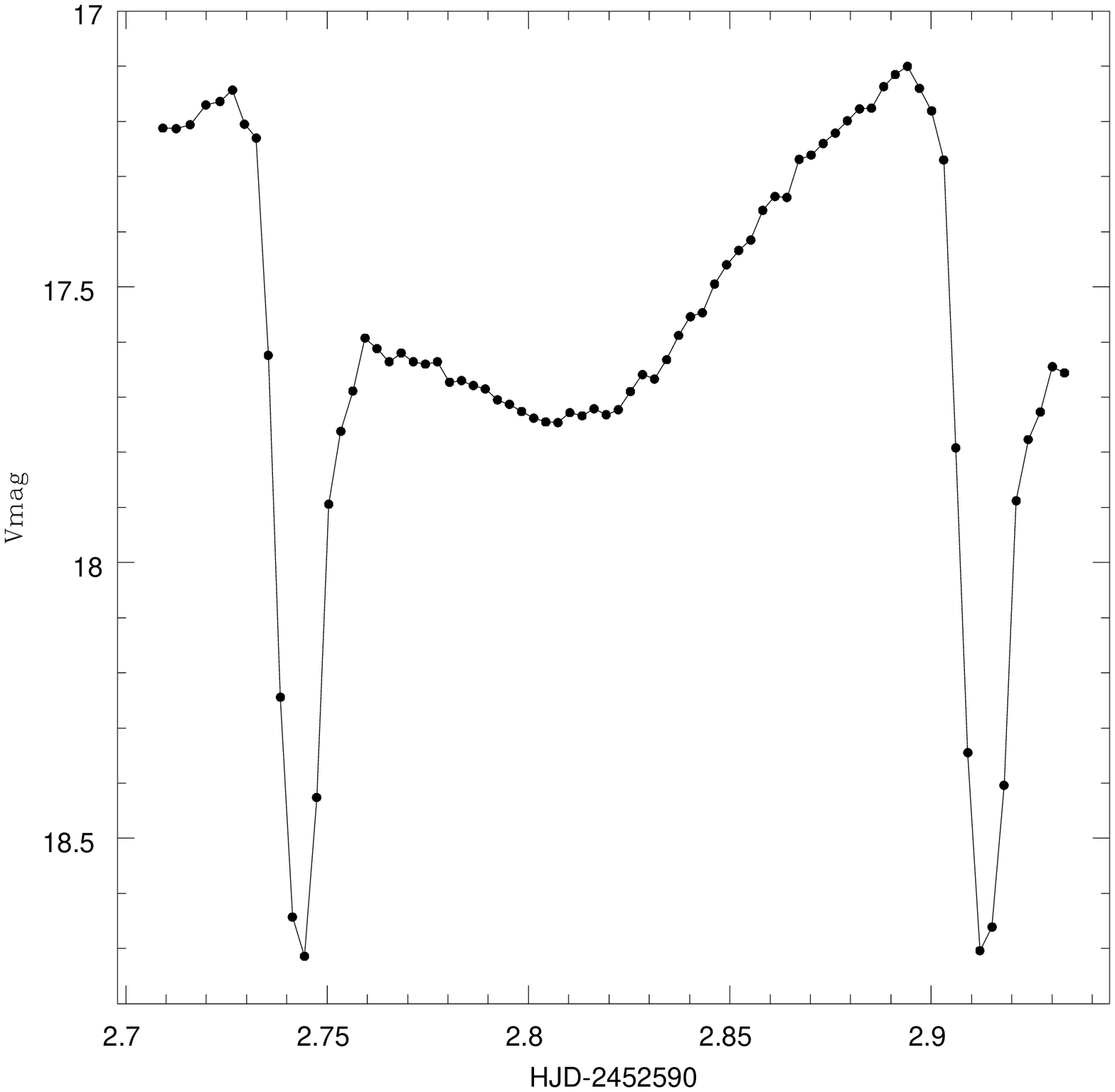}
\figcaption{NOFS light curve of SDSS0407, showing the deep eclipse and prominent orbital hump.}
\end{figure}
\begin{figure}
\plotone{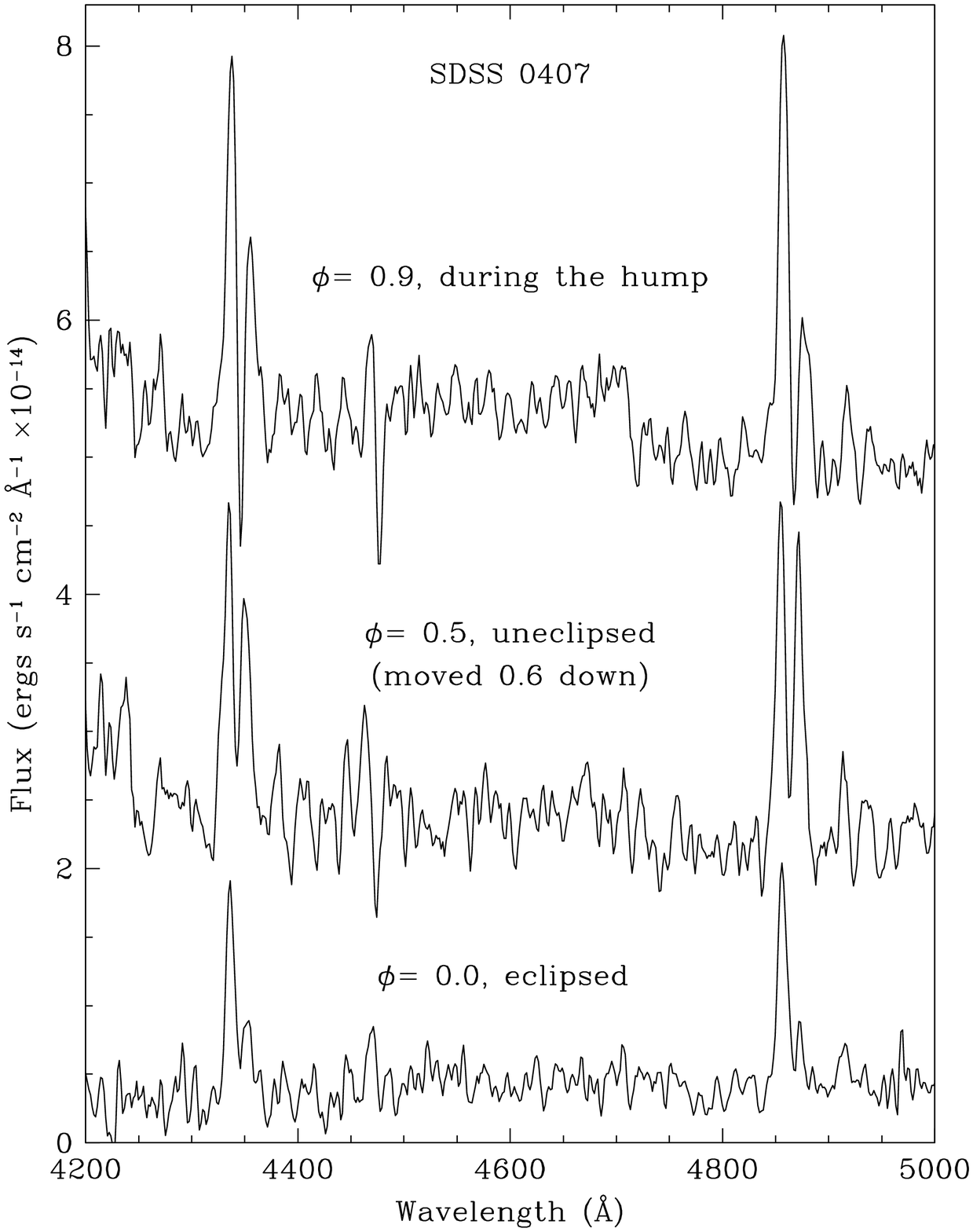}
\figcaption{APO spectra showing the eclipse, hump, and uneclipsed phases of SDSS0407.}
\end{figure}
\clearpage
\begin{figure}
\plotone{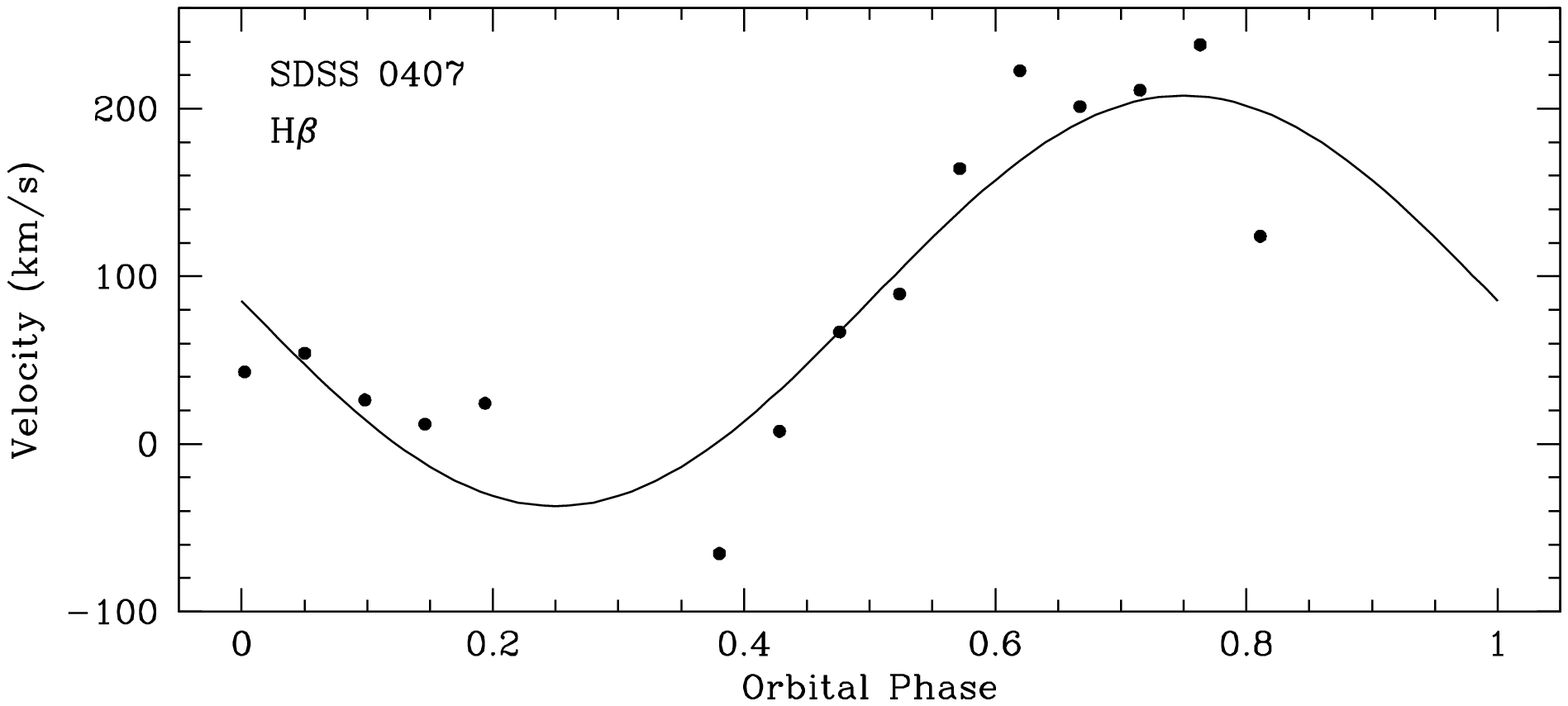}
\figcaption{Velocity curve of SDSS0407 with the best fit sinusoid.}
\end{figure}
\begin{figure}
\epsscale{0.8}
\plotone{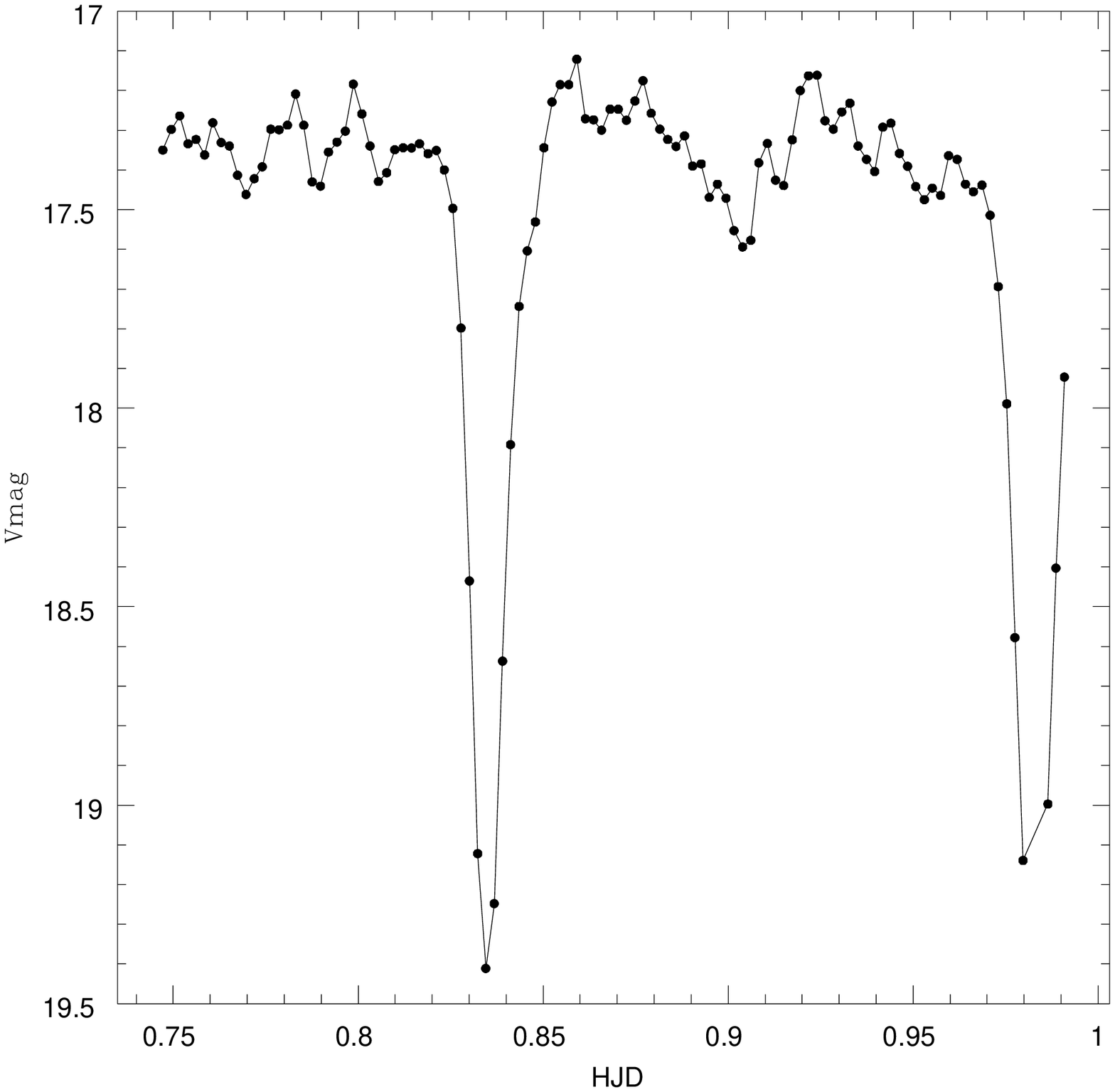}
\figcaption{NOFS light curve of SDSS0920, showing its deep eclipses.}
\end{figure}

\begin{figure}
\epsscale{1.0}
\plotone{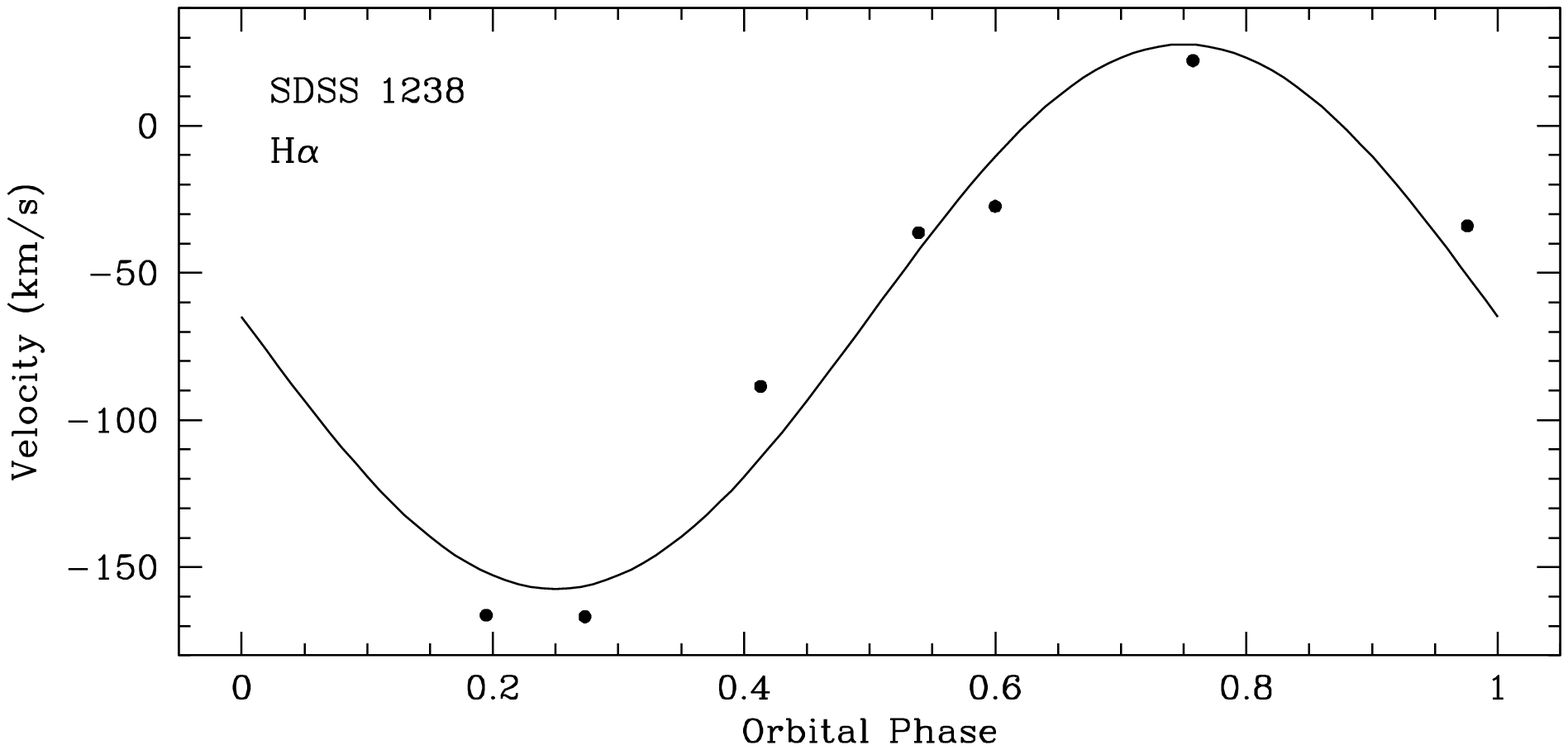}
\figcaption{Velocity curve of SDSS1238 with the best fit sinusoid.}
\end{figure}

\begin{figure}
\plotone{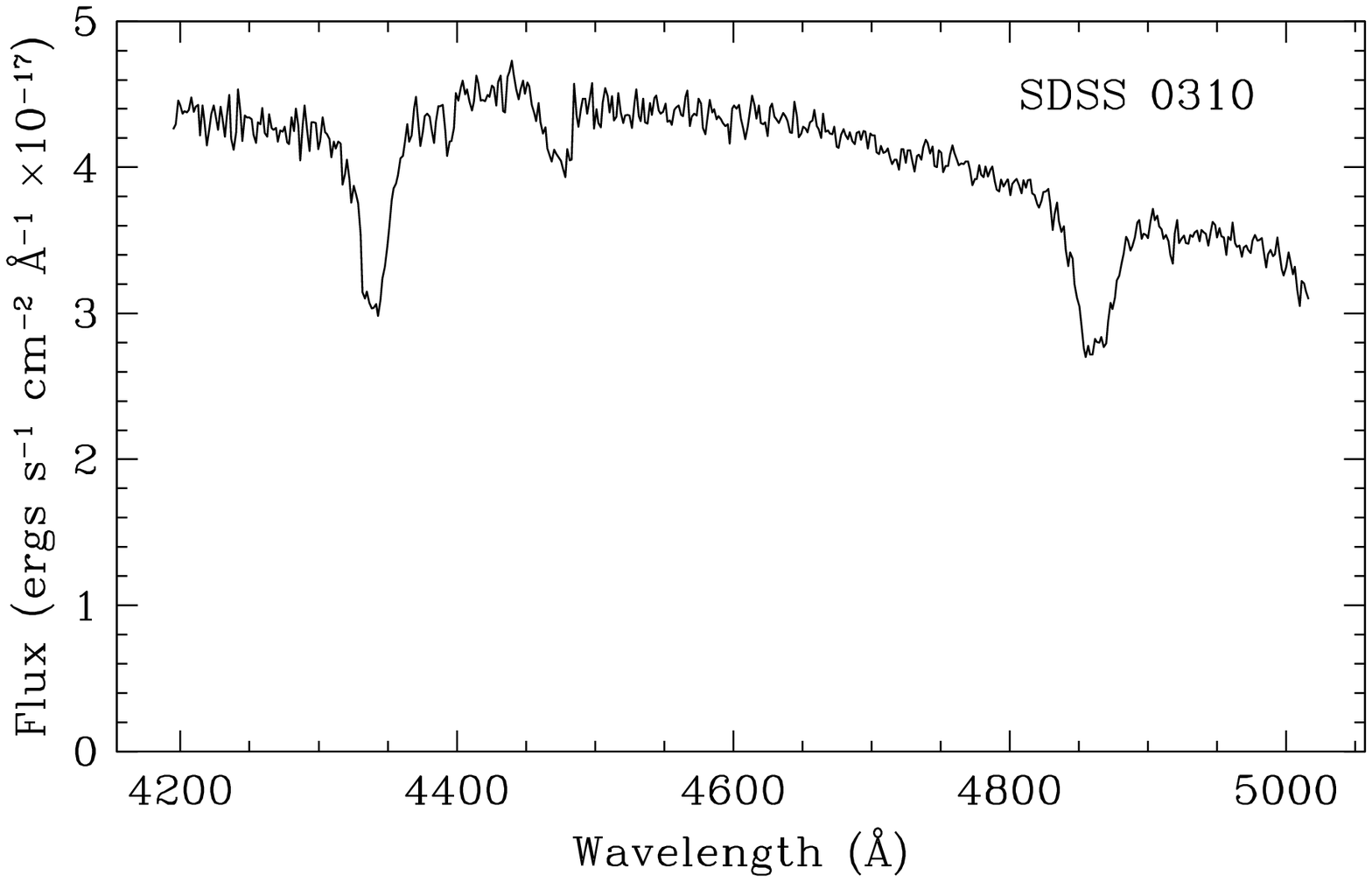}
\figcaption{APO outburst spectrum of SDSS0310.}
\end{figure}
\begin{figure}
\plotone{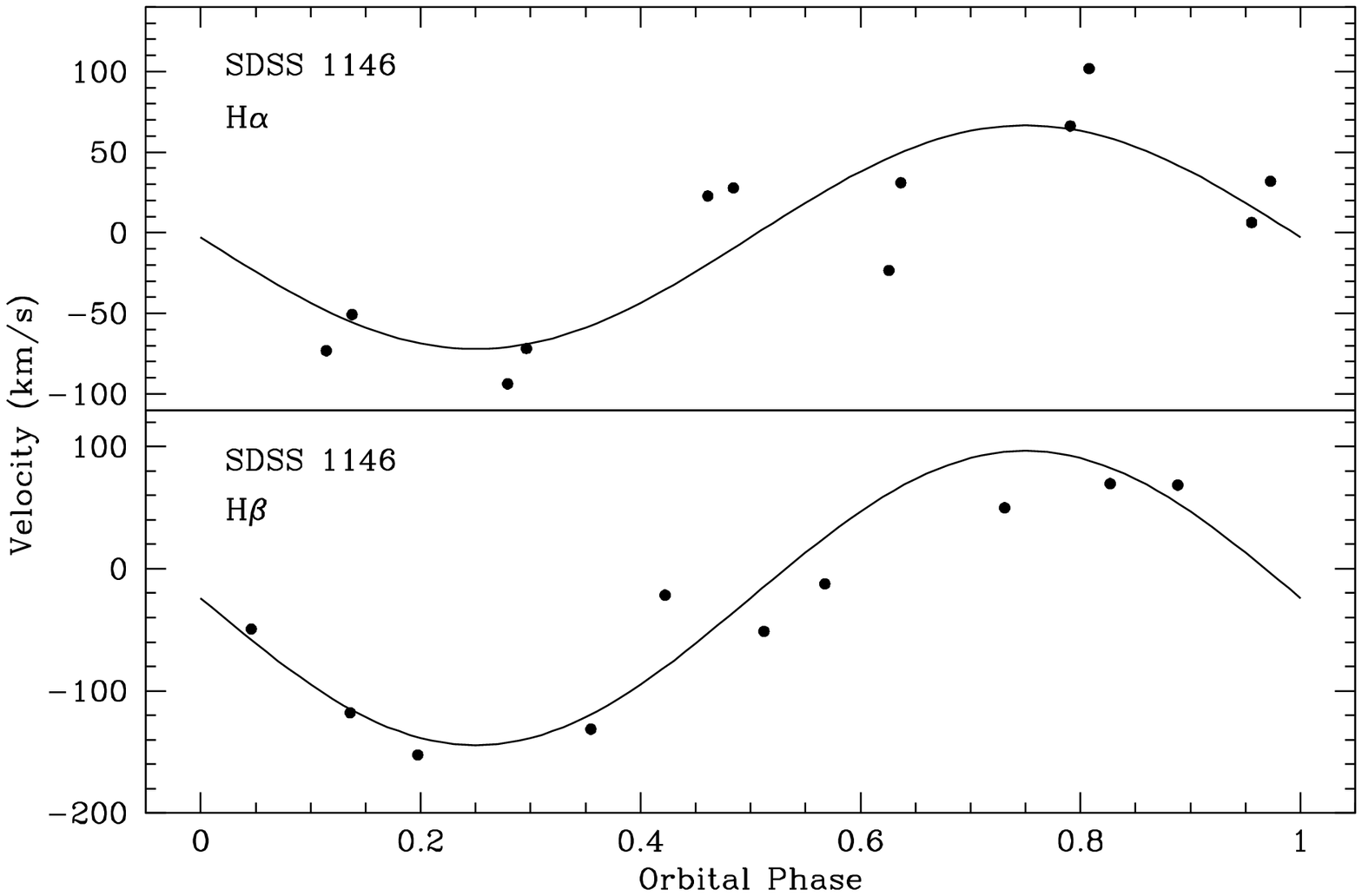}
\figcaption{Velocity curves of SDSS1146 with the best fit sinusoids superposed.}\end{figure}
\begin{figure}
\plotone{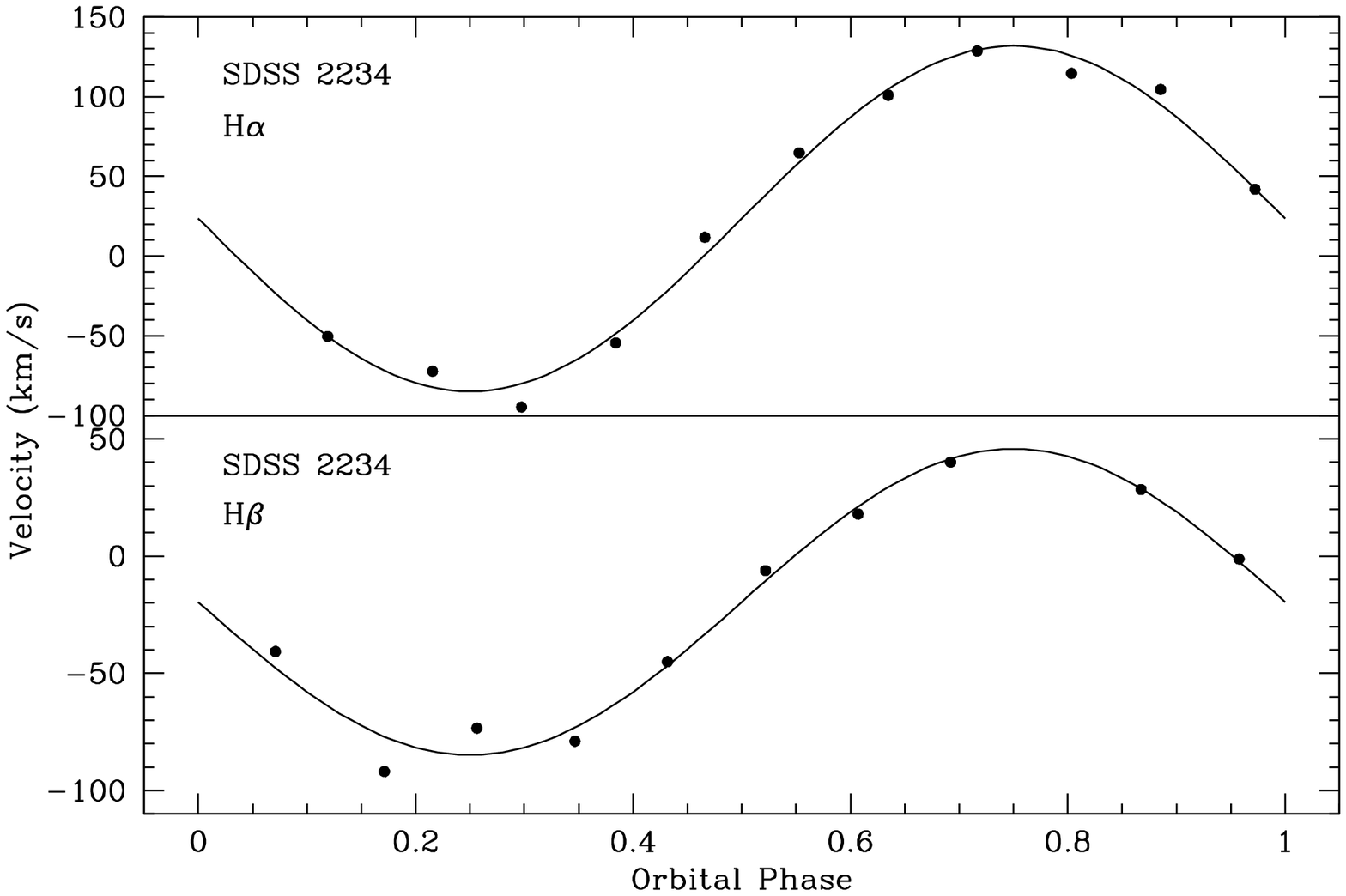}
\figcaption{Velocity curves of SDSS2234 with the best fit sinusoids superposed.}\end{figure}
\begin{figure}
\plotone{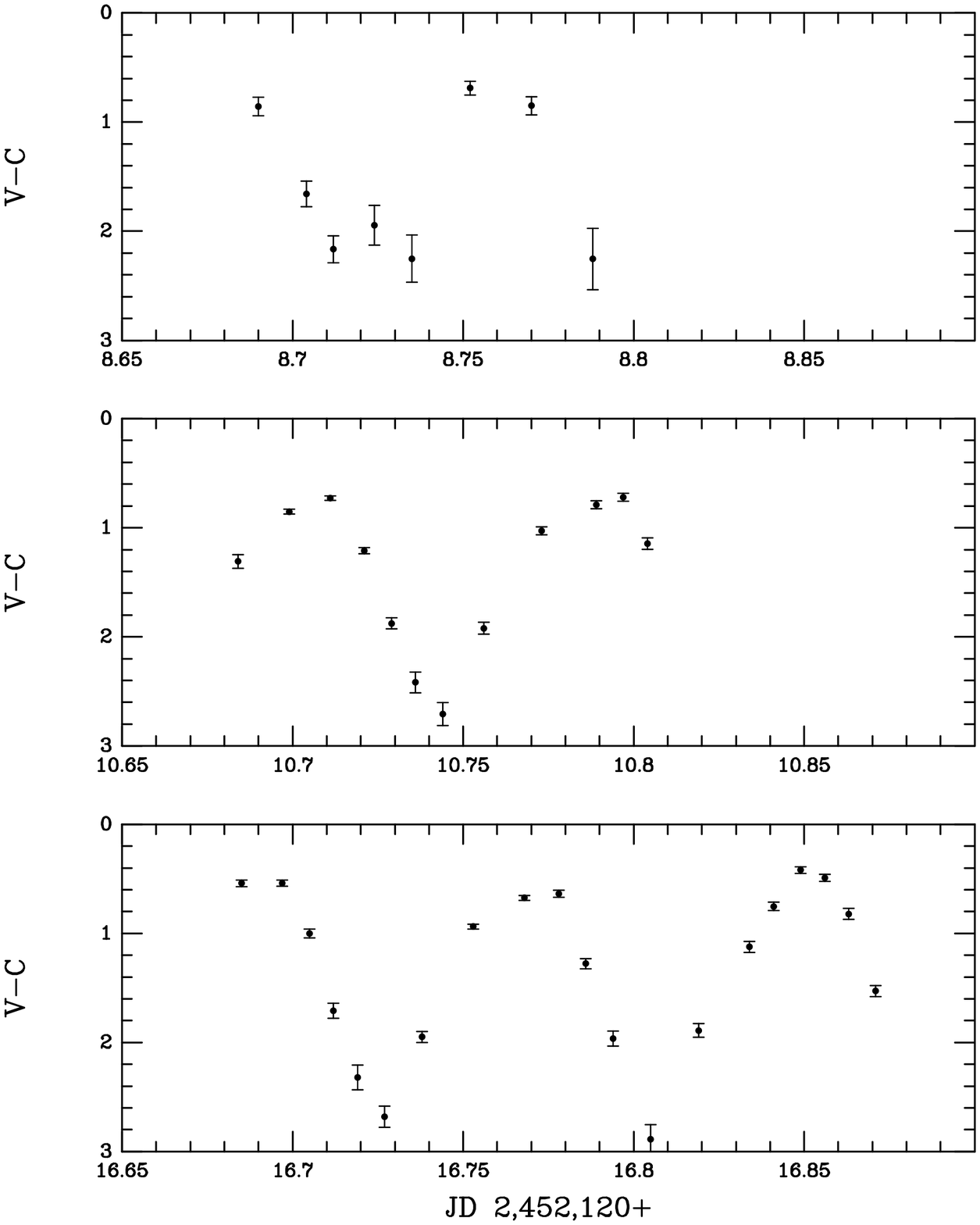}
\figcaption{MRO differential light curves of SDSS1700.}
\end{figure}
\begin{figure}
\plotone{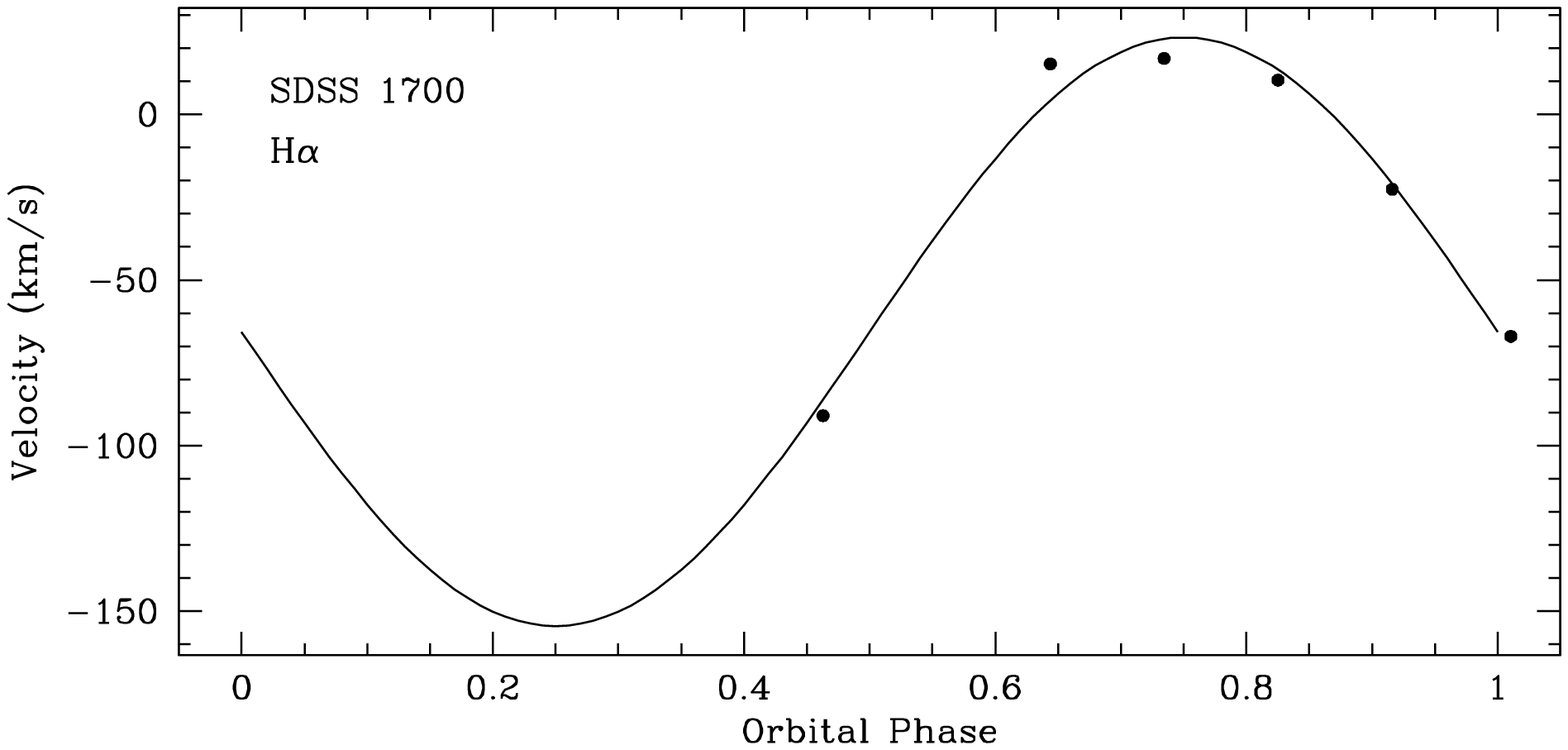}
\figcaption{Velocity curve of SDSS1700 with the best fit sinusoid found by using the photometric period.}
\end{figure}

\begin{figure}
\epsscale{1.0}
\plotone{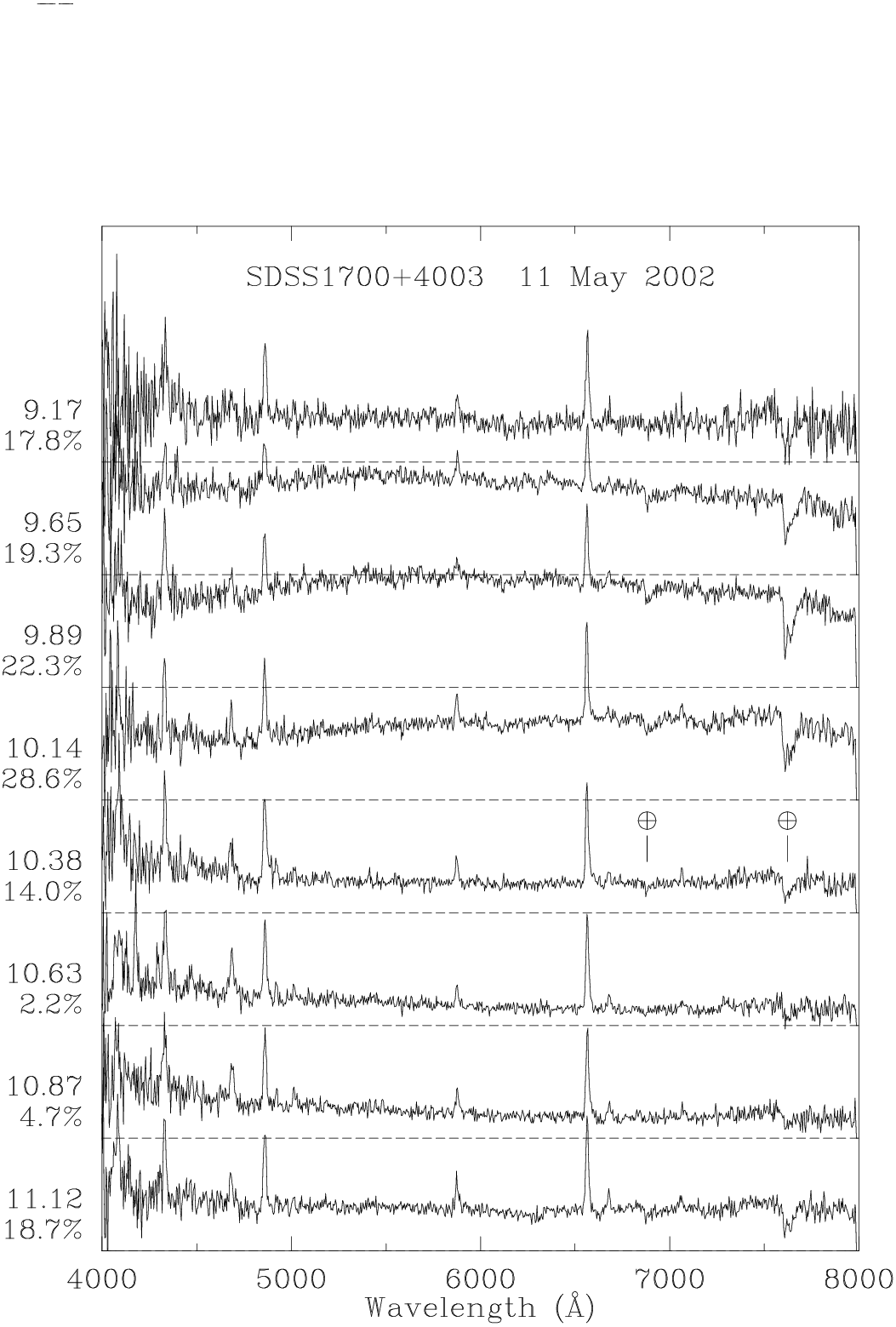}
\figcaption{Sequence of low-resolution ($\sim$13\AA) spectroscopy covering one
period of SDSS1700.  The UT and broadband ($\lambda\lambda4500-7500$\AA)
circular polarization are listed for each panel at the left.}
\end{figure}
\begin{figure}
\epsscale{1.0}
\resizebox*{0.9\textwidth}{!}{\rotatebox{-90}{\plotone{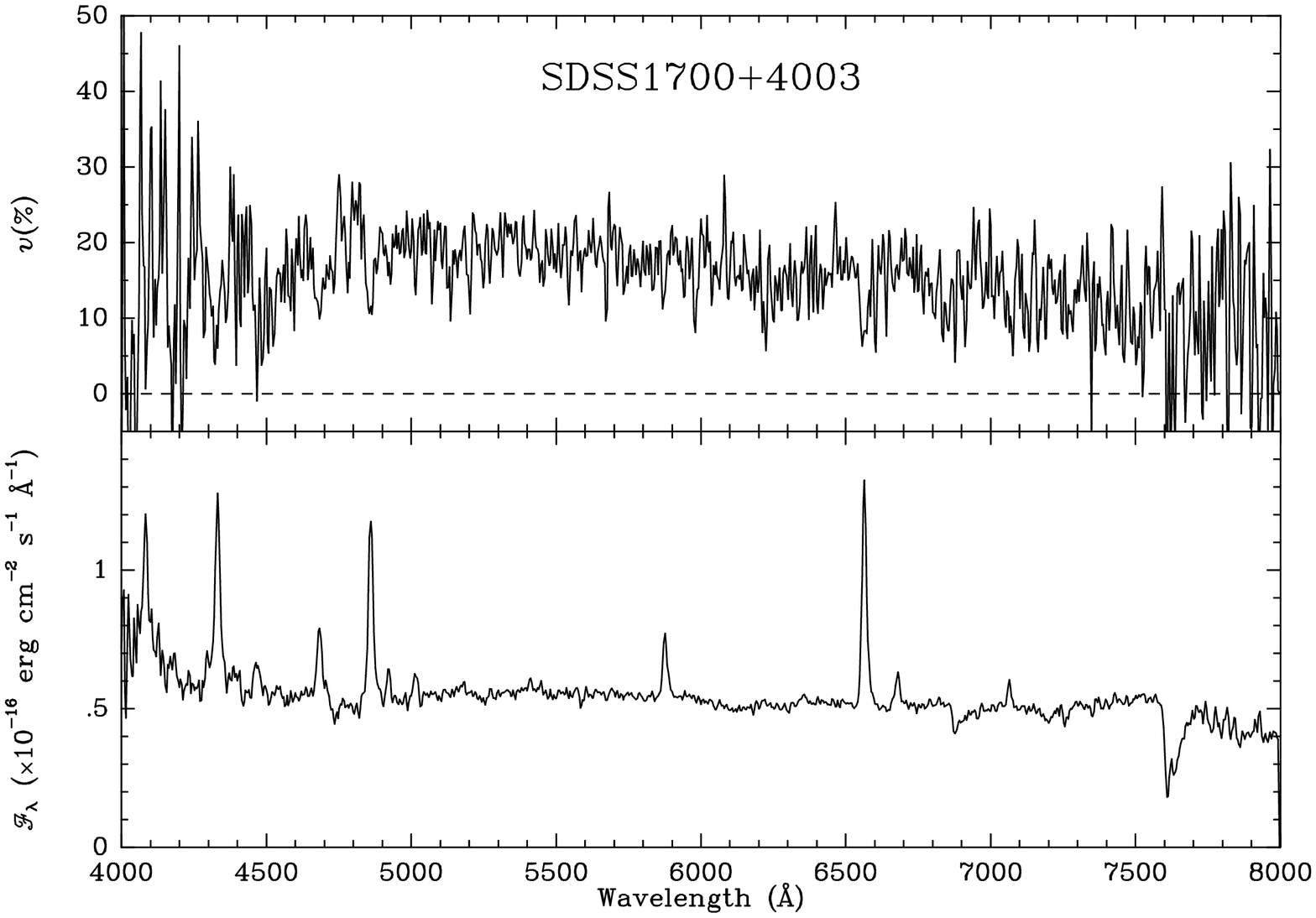}}}
\figcaption{Coadded circular polarization {\it (top)\/} and spectral flux {\it 
(bottom)\/} for SDSS1700.}
\end{figure}
\begin{figure}
\plotone{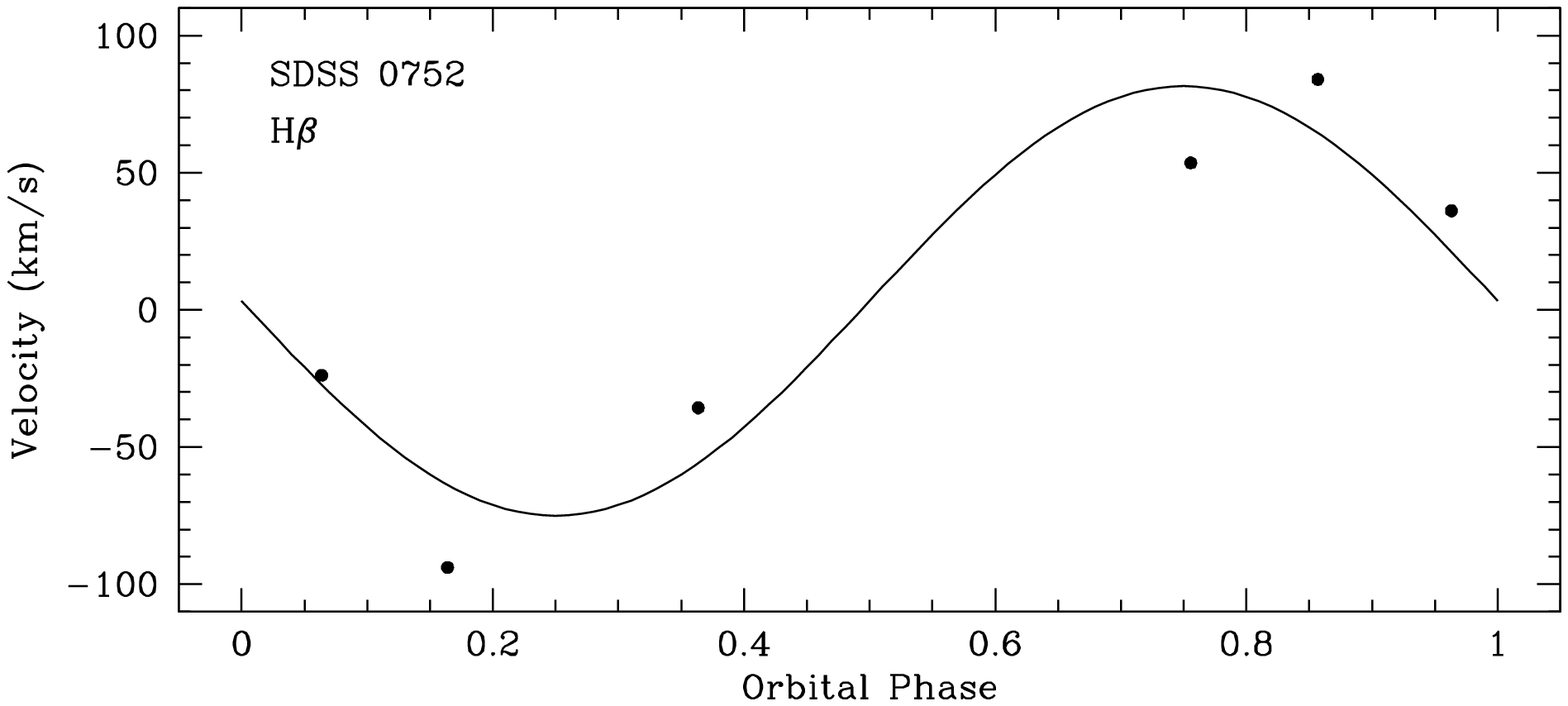}
\figcaption{Velocity curve of SDSS0752 with the best fit sinusoid.}
\end{figure}
\begin{figure}
\plotone{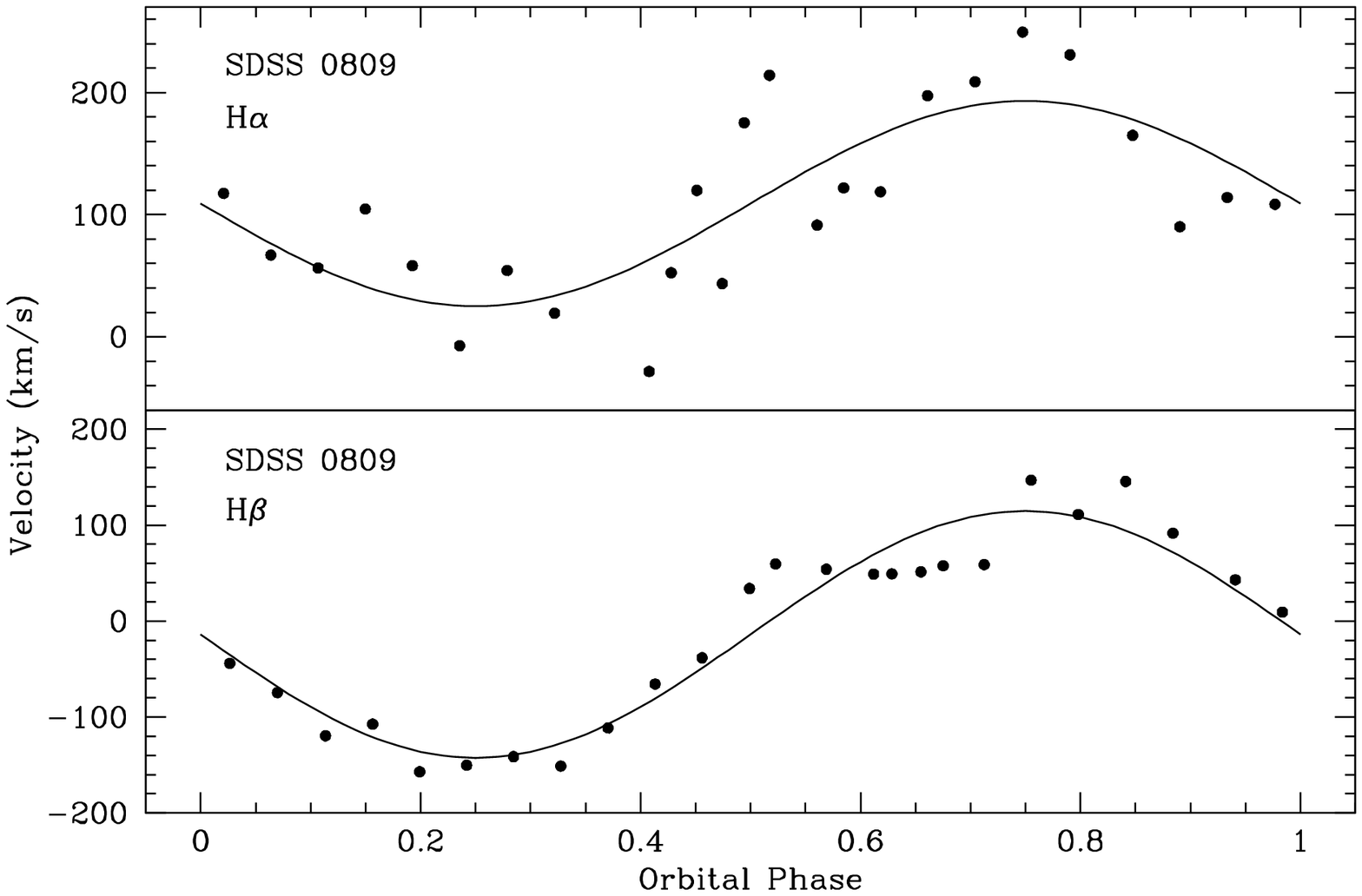}
\figcaption{Velocity curves of SDSS0809 with the best fit sinusoids.}
\end{figure}
\begin{figure}
\plotone{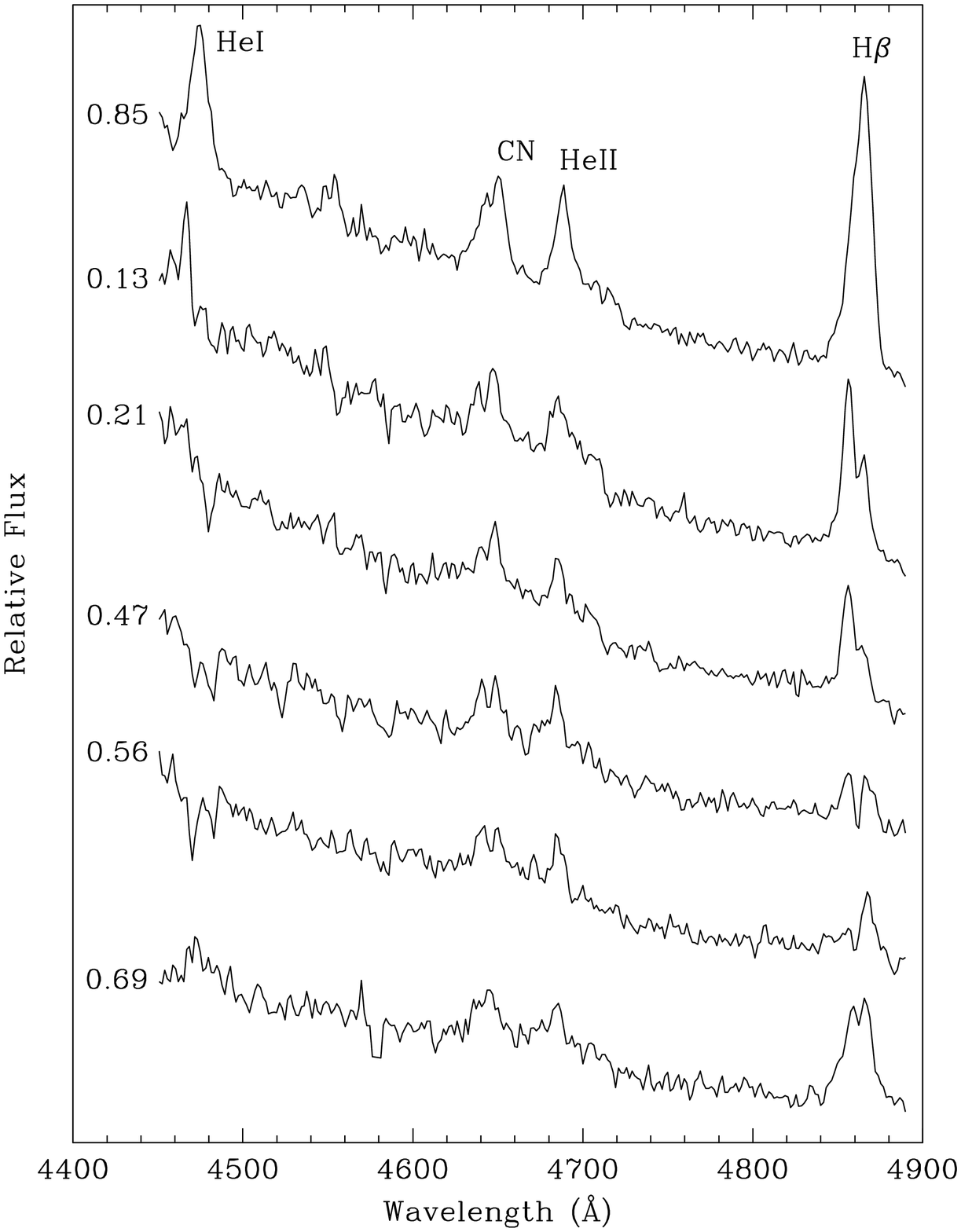}
\figcaption{Time-resolved spectra of SDSS0809 showing the changing line structure and
deep absorption apparent in the Balmer and \ion{He}{1} lines during phases 0.21-0.56.
Flux scale is in F$_{\lambda}$ units with each successive spectrum offset for
clarit`y.}
\end{figure}
\begin{figure}
\plotone{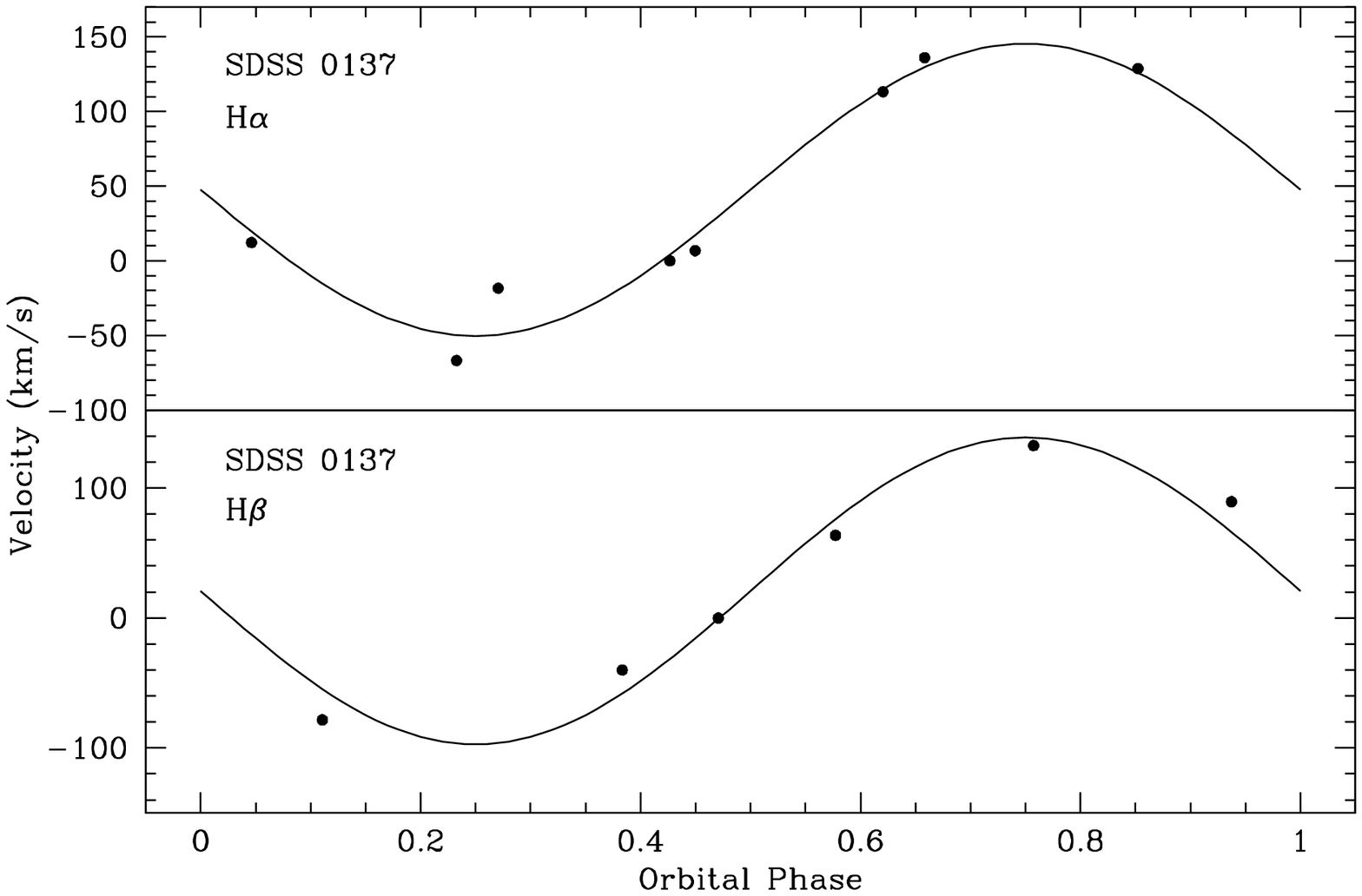}
\figcaption{Velocity curves of SDSS0137 with the best fit sinusoids.}
\end{figure}

\end{document}